\documentclass[prl,aps,superscriptaddress,footinbib,twocolumn]{revtex4-2}

\usepackage[latin9]{inputenc}
\usepackage{xcolor}
\usepackage{amsmath}
\usepackage{amssymb}
\usepackage{stmaryrd}
\usepackage{graphicx}
\usepackage{scalerel}
\usepackage{adjustbox}
\usepackage{tikz}
\usepackage{enumitem}
\usepackage[unicode=true,
bookmarks=true,bookmarksnumbered=false,bookmarksopen=false,
breaklinks=false,pdfborder={0 0 1},backref=false,colorlinks=true]
{hyperref}
\hypersetup{
	linkcolor=magenta, urlcolor=blue, citecolor=blue, pdfstartview={FitH}, hyperfootnotes=false, unicode=true}

\makeatletter
\usepackage{amsfonts}
\usepackage{tabularx}
\usepackage{booktabs}
\usepackage{multirow}
\usepackage{bm}
\usepackage{diagbox}
\usepackage{times}
\usepackage{cellspace}
\makeatother

\begin{document}
\title{Proving olympiad geometry theorems on a superconducting quantum processor}
  
\affiliation{School of Physics, ZJU-Hangzhou Global Scientific and Technological Innovation Center, \\and Zhejiang Key Laboratory of Micro-nano Quantum Chips and Quantum Control, Zhejiang University, Hangzhou, China\\
$^2$ Center for Quantum Information, IIIS, Tsinghua University, Beijing 100084, China\\
$^{3}$ Hefei National Laboratory, Hefei 230088, China\\
$^{4}$ Shanghai Qi Zhi Institute, Shanghai 200232, China}
  
\author{Ning Wang$^{1}$}\thanks{These authors contributed equally to this work.}
\author{Zheng-Zhi Sun$^{2}$}\thanks{These authors contributed equally to this work.}
\author{Zhengyi Cui$^{1}$}	
\author{Yiren Zou$^{1}$}
\author{Aosai Zhang$^{1}$}	
\author{Fanhao Shen$^{1}$}
\author{Jiarun Zhong$^{1}$}
\author{Zehang Bao$^{1}$}
\author{Zitian Zhu$^{1}$}
\author{Han Wang$^{1}$}
\author{Jia-Nan Yang$^{1}$}
\author{Jiayuan Shen$^{1}$}
\author{Gongyu Liu$^{1}$}
\author{Yanzhe Wang$^{1}$}
\author{Yihang Han$^{1}$}
\author{Yiyang He$^{1}$}
\author{Jiahua Huang$^{1}$}
\author{Sailang Zhou$^{1}$}
\author{Xinrong Zhang$^{1}$}
\author{Yaozu Wu$^{1}$}
\author{Zixuan Song$^{1}$}
\author{Jinfeng Deng$^{1}$}
\author{Hang Dong$^{1}$}
\author{Qi Ye$^{2}$}
\author{Weikang Li$^{2}$}
\author{Si Jiang$^{2}$}
\author{Yixuan Ma$^{2}$}
\author{Shuangyue Geng$^{2}$}
\author{Zhide Lu$^{4}$}
\author{Chao Song$^{1,3}$} \email{chaosong@zju.edu.cn}
\author{Hekang Li$^{1,3}$}
\author{Pengfei Zhang$^{1,3}$}
\author{Qiujiang Guo$^{1,3}$}
\author{H. Wang$^{1, 3}$}\email{hhwang@zju.edu.cn}
\author{Dong-Ling Deng$^{2, 4, 3}$}\email{dldeng@tsinghua.edu.cn}

\begin{abstract}\noindent
\textbf{Automated theorem proving seeks to use computational systems to prove or disprove mathematical and logical statements \cite{Fitting2012First,Loveland2016Automated}. It underpins a wide range of applications, and enhancing theorem-proving capabilities remains a central objective in artificial intelligence \cite{Hasan2015Formal}. Although recent neuro-symbolic systems have achieved remarkable progress \cite{Trinh2024Solving,Guo2025DeepSeek,Hubert2026Olympiad,Gowers2025Conjecture}, their operation is ultimately constrained by classical computational architectures. Quantum computing \cite{Nielsen201206Quantum}, by contrast, enables information encoding and coherent parallelism beyond classical limits \cite{Arute2019Quantum,Wu2021Strong,Madsen2022Quantum,Zhong202012Quantum,Google2025Observation,King2025Beyond}, raising the possibility of accelerating structured symbolic deduction \cite{Sun2026Quantum}. Here we report the experimental realization of automated geometry theorem proving on a fully programmable superconducting quantum processor. We develop two complementary quantum proving frameworks. 
The first implements Wu's algebraic elimination method using quantum pseudo-division, with multivariate polynomials represented in superposition states, enabling quantum algebraic theorem proving. The second implements the full-angle method as backward symbolic reasoning through a hybrid quantum strategy-guided architecture, demonstrating a general route toward quantum symbolic proof search.
As illustrative examples, we prove two theorems on a superconducting quantum processor: the perpendicularity of the diagonals of a square and a 1978 International Mathematical Olympiad geometry problem. Our results establish, at the experimental level, automated logical reasoning as a viable task for near-term quantum processors and provide a concrete pathway toward quantum-enhanced symbolic intelligence.} 
\end{abstract}
	
\maketitle

\noindent Automated theorem proving  is a central challenge in artificial intelligence and a crucial tool for formal verification \cite{Hasan2015Formal}. Geometry provides an intuitive yet demanding setting: its statements are readily visualized, but their proofs require precise representations and rigorous chains of deduction. Wu's method translates geometric relations into polynomial equations and establishes conclusions by algebraic elimination under suitable non-degeneracy conditions \cite{Wu1978Decision}. The full-angle method instead constructs readable proofs by manipulating relations between angles \cite{Chou199612Automated}. Recent neuro-symbolic systems, such as AlphaGeometry \cite{Trinh2024Solving} and TongGeometry \cite{Zhang2026Proposing}, combine neural guidance with symbolic deduction to solve challenging International Mathematical Olympiad (IMO) geometry problems~\cite{Trinh2024Solving,Zhang2026Proposing}, alongside broader advances in learning-based mathematical reasoning~\cite{Guo2025DeepSeek,Hubert2026Olympiad}. Despite these advances, automated theorem proving in geometry may still face a fundamental scalability barrier: the size of intermediate expressions and the number of candidate inference steps can grow rapidly with problem complexity, imposing unattainable time and memory costs on classical computers.

Quantum computing, grounded in the principles of quantum mechanics, provides a fundamentally different substrate for information processing \cite{Nielsen201206Quantum}. By leveraging peculiar quantum features, such as superposition and entanglement, quantum processors would manipulate information in parallel in exponentially large Hilbert spaces, allowing unparalleled opportunity for addressing the scalability challenge in classical automated theorem proving. 
In recent years, exciting progress has been achieved across multiple quantum hardware platforms, including programmable superconducting circuits \cite{Arute2019Quantum, Wu2021Strong, Krinner202205Realizing, Kim202306Evidence,Xu202407Non,Jin2025Topological}, trapped ions \cite{Egan202110Fault, Moses202312Race}, photonic systems \cite{Zhong202012Quantum, Madsen2022Quantum}, and Rydberg-atom arrays \cite{Ebadi202107Quantum,Bluvstein2024Logical}. These advances, together with developments in experiments exhibiting quantum computational advantage \cite{Arute2019Quantum,Zhong202012Quantum,Wu2021Strong,Madsen2022Quantum}  and demonstrating quantum error correction \cite{Acharya202302Suppressing,Sivak202303Real,Ni202303Beating,Bluvstein2024Logical,Gupta2024Encoding,Acharya2025, Wang2026Demonstration,Lacroix2025Scaling}, are steadily pushing quantum devices toward regimes where increasingly complex and noise-resilient computations become feasible. Parallel to these developments, the intersection of quantum computing and machine learning has given rise to a new research frontier of quantum machine learning \cite{Biamonte2017Quantum,Dunjko2018Machine,DasSarma2019Machine,Cerezo2022Challenges,Li2025Pitfalls}. A variety of quantum learning algorithms with potential advantages have been proposed \cite{Harrow2009Quantum,Lloyd2014Quantum,Dunjko2016Quantum,Gao2018Quantum,Liu2021Rigorous,Jerbi2023Quantum}  and some of them have been demonstrated in proof-of-principle experiments  \cite{Havlicek2019Supervised,Hu2019Quantum,Huang2022Quantum,Ren2022Experimental}. More recently, a theoretical framework for quantum automated theorem proving has been introduced, extending quantum computation beyond predominantly numerical tasks to structured symbolic reasoning~\cite{Sun2026Quantum}. Within this framework, generic reasoning problems admit a quadratic quantum speedup.  Yet, experimental demonstration of quantum automated theorem proving remains a notable challenge and has not been reported so far.

\begin{figure*}[htbp]
\centering
\includegraphics[width=\textwidth]{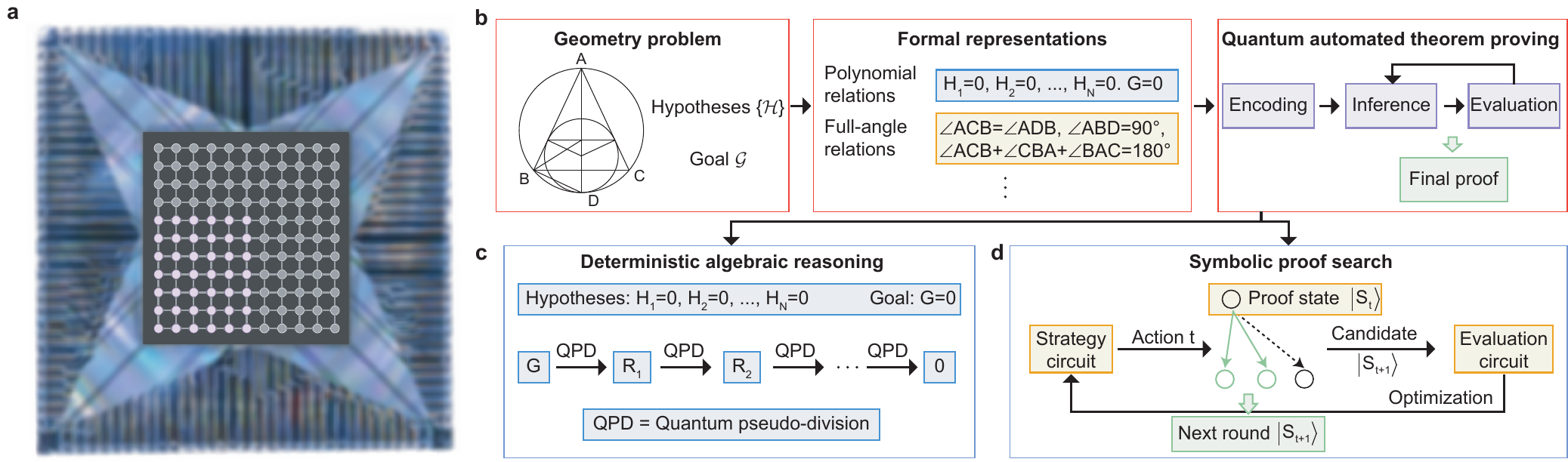}
\caption{\textbf{The quantum processor and overview of quantum automated theorem proving. }
\textbf{a}, Superconducting quantum processor used in the experiments. The qubits marked in purple are selected to implement the quantum reasoning circuits.
\textbf{b}, Overview of quantum automated theorem proving. Automated theorem proving seeks to derive a target conclusion $\mathcal{G}$ from a set of hypotheses $\mathcal{H}$ through formally valid inference. The geometry configuration shown is taken from a problem of the 1978 IMO and serves as an illustrative example. The hypotheses and conclusion are translated into exact formal representations, including polynomial equations and full-angle relations, and subsequently encoded into quantum registers. Quantum circuits implement inference and evaluation, with the outcome classified as proved, disproved or undetermined according to the corresponding termination criterion.
\textbf{c}, Deterministic algebraic reasoning based on Wu's method. The hypotheses are represented by the polynomial knowledge base $H_1=0,H_2=0,\ldots,H_N=0$, and the conclusion by the goal polynomial $G$. Successive quantum pseudo-division (QPD) operations reduce $G$ through a sequence of pseudo-remainders, with a vanishing final pseudo-remainder establishing the conclusion (Supplementary Information Sec.1).
\textbf{d}, Symbolic proof search. The strategy circuit generates an inference action $A_t$ that transforms the proof state $\lvert S_t\rangle$ into a candidate output $\lvert S_{t+1}\rangle$. The action and candidate output are encoded as coherent superpositions of inference operations and proof states, respectively. We use an evaluation circuit to estimate the value function for the candidate output and iteratively optimize the strategy circuit. The resulting state $\lvert S_{t+1}\rangle$ is used in the next reasoning round (see Supplementary Information Sec.2 for an in-depth discussion).}
\label{fig-Overview}
\end{figure*}

Here we report an experimental realization of quantum automated theorem proving on a programmable superconducting quantum processor (Fig.~\ref{fig-Overview}\textbf{a}), using geometry theorems serving as representative examples. We experimentally realize two complementary forms of quantum reasoning, spanning algebraic and symbolic deduction. For algebraic deduction, we map the pseudo-division procedure of Wu's method onto quantum circuits and demonstrate the complete proof workflow using the perpendicularity of the diagonals of a square as a benchmark. For symbolic deduction, we combine the full-angle representation with a hybrid quantum-classical proof-search architecture. A parameterized quantum circuit proposes inference actions, while a fixed reasoning circuit implements the underlying formal rules. The resulting procedure constructs a human-readable proof of a geometry problem from the 1978 IMO. Together, these experiments demonstrate that programmable quantum processors can carry out structured formal reasoning and establish an experimental basis for quantum automated theorem proving and symbolic intelligence.

\begin{figure*}[htbp]
    \centering
    \includegraphics[width=\textwidth]{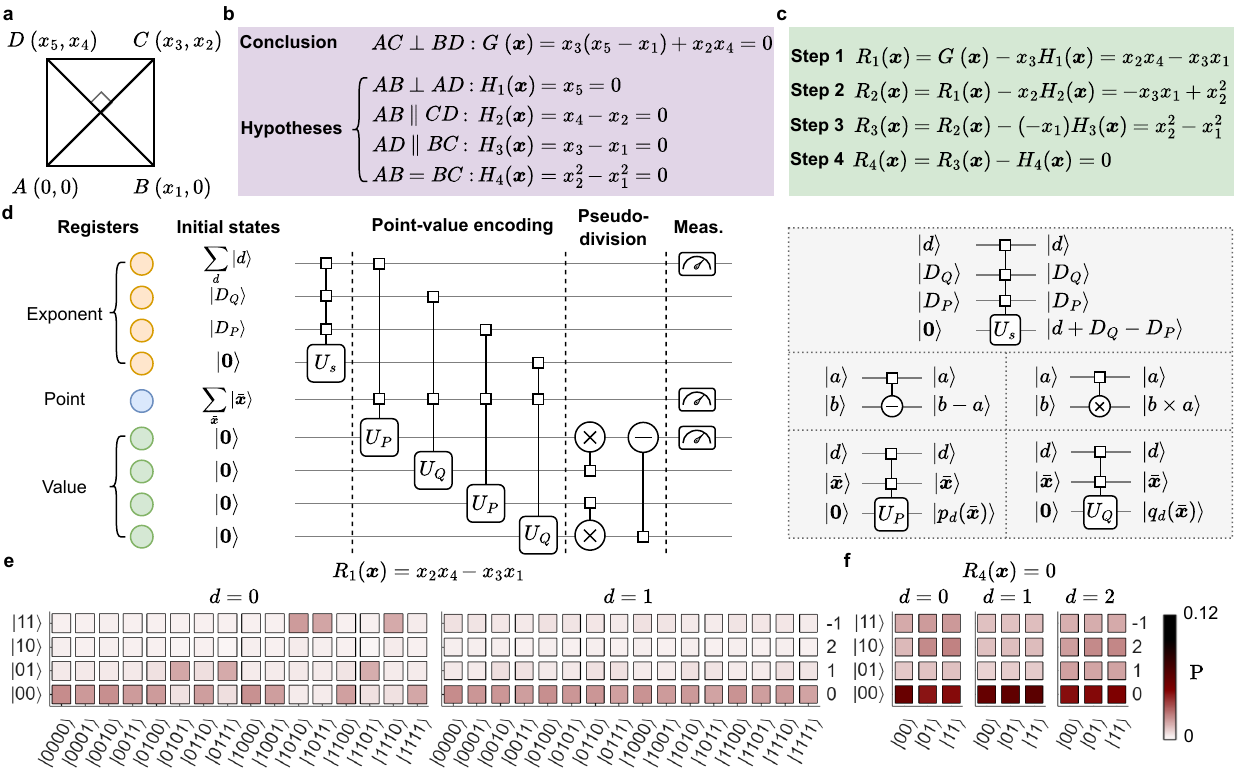}
    \caption{
    \textbf{Demonstration of quantum algebraic theorem proving with a square geometry problem}. 
    \textbf{a}, Coordinate representation of the square $ABCD$, with $A=(0,0)$, $B=(x_1,0)$, $C=(x_3,x_2)$ and $D=(x_5,x_4)$.
    \textbf{b}, Polynomial formulation of the theorem. The geometric hypotheses $AB\perp AD$, $AB\parallel CD$, $AD\parallel BC$ and $AB=BC$ are represented by the polynomials $H_1(\boldsymbol{x}),\ldots,H_4(\boldsymbol{x})$, and the conclusion $AC\perp BD$ by the polynomial $G(\boldsymbol{x})$, where $\boldsymbol{x}=(x_1,x_2,\ldots,x_5)$.
    \textbf{c}, Algebraic proof by successive pseudo-division. The variables $x_5$, $x_4$, $x_3$ and $x_2$ are eliminated sequentially using $H_1$, $H_2$, $H_3$ and $H_4$, respectively. The resulting pseudo-remainder chain terminates at $R_4(\boldsymbol{x})=0$, establishing the conclusion.
    \textbf{d}, The conceptual quantum circuit for dividing polynomial $P$ by $Q$. We partition all nine registers into three groups for storing different information of the two polynomials and have already initialized their states as annotated. The boldface label $|\mathbf{0}\rangle$ denotes the all-zero state of the corresponding register, which may comprise multiple qubits. At the beginning, we modify the fourth exponent register by controlled $U_s$ for the subsequent second controlled $U_Q$ operation. Then we write four value registers by controlled unitary $U_P$ or $U_Q$ based on the shared point and corresponding exponent registers, and apply quantum arithmetic operations on two of them. Lastly, we measure the necessary registers to reconstruct the resulting polynomial. In the right gray box, we show the functions of all controlled unitary gates involved in the quantum circuit.
    \textbf{e-f}, The experimental point-value correspondence of the first and final step. As shown in \textbf{e} (\textbf{f}), these results are classified into two (three) sectors depending on the value of $d$ and the x-axis labels the joint states of point qubits $|x_1x_2x_3x_4\rangle$ ($|x_1\rangle$). The joint states of the measured value register are depicted on the left side, with their corresponding encoded values shown on the right side. Colors indicate the probability obtained from 3000 runs of measurements, with each run taking 3000 shots.} \label{fig-Square}
\end{figure*}

\vspace{.5cm}
\noindent\textbf{\large{Framework and experimental setup}}

\noindent Quantum automated theorem proving addresses problems defined by a set of hypotheses $\mathcal{H}=\{\mathcal{H}_1,\mathcal{H}_2,\ldots\}$ and a target conclusion $\mathcal{G}$ (Fig.~\ref{fig-Overview}\textbf{b}). The mathematical statements are first cast into exact formal representations suitable for quantum processing. The resulting knowledge, proof states and inference operations are encoded in quantum registers, allowing formal reasoning and evaluation to be implemented through quantum circuits. Depending on the reasoning procedure and its termination criterion, the computation may prove the conclusion, disprove it or return an undetermined outcome \cite{Sun2026Quantum}.

This framework supports two complementary approaches. In algebraic reasoning, mathematical statements that admit polynomial formulations are processed through quantum polynomial operations, with the final result certified by an algebraic termination criterion (Fig.~\ref{fig-Overview}\textbf{c}). In strategy-guided symbolic proof search, a quantum knowledge base and the current proof state are processed by strategy, reasoning and evaluation circuits to identify a sequence of formally valid inference steps (Fig.~\ref{fig-Overview}\textbf{d}). The resulting proof state can be passed directly to the subsequent reasoning step without extracting and re-encoding the intermediate statement. More generally, retaining knowledge and proof states in quantum form provides a route towards coherent proof search, in which superpositions of candidate actions and reasoning paths could be maintained and processed across multiple inference steps. The detailed representations and circuit constructions are introduced in the following sections.

Our experiments are performed on a two-dimensional (2D) flip-chip superconducting quantum processor comprising 121 frequency-tunable transmon qubits arranged in an $11\times11$ square lattice (Fig.~\ref{fig-Overview}\textbf{a}). Nearest-neighbour interactions are mediated by 220 tunable couplers, and the processor supports arbitrary single-qubit rotations, two-qubit controlled-phase gates and frequency-multiplexed dispersive readout. Connected subsets of qubits are selected to implement the reasoning circuits, which are compiled into the native gate set with multi-qubit operations and qubit layouts optimized to reduce circuit depth. The fidelity performance of simultaneous single-qubit gates, two-qubit gates, and measurement gates are better than $99.9\%$, $99\%$, and $98\%$, respectively. 
More detailed device parameters, calibration procedures and compilation methods are provided in the Supplementary Information Sec.3.

\vspace{.5cm}
\noindent\textbf{\large{Quantum algebraic theorem proving}}

\noindent We present quantum algebraic theorem proving by mapping polynomial deduction onto a sequence of quantum pseudo-divisions (QPD) \cite{Sun2026Quantum}. We implement Wu's method on the superconducting quantum processor, using a proof of the perpendicularity of the diagonals of a square as an example (Fig.~\ref{fig-Square}\textbf{a}). The coordinates of four corner points $A$-$D$ are parameterized by the variables $\boldsymbol{x}=(x_1,x_2,\ldots,x_5)$, and the geometric hypotheses together with the target conclusion are translated into multivariate polynomial equations (Fig.~\ref{fig-Square}\textbf{b}). The proof comprises four QPD steps, each of which reduces the current polynomial using one of the hypothesis polynomials and eliminates one variable (Fig.~\ref{fig-Square}\textbf{c}). The vanishing of the final pseudo-remainder establishes the conclusion \cite{Sun2026Quantum}.

We encode each polynomial in the point-value form \cite{Sun2026Quantum}. For a polynomial $P$ whose leading variable is to be eliminated, the coefficient associated with each power of that variable is regarded as a polynomial in the remaining variables and evaluated at a set of sampling points sufficient to determine it uniquely. The exponent, sampling point and corresponding coefficient value are stored in separate quantum registers, with a controlled unitary $U_P$ writing the coefficient value according to the exponent and point inputs. The divisor polynomial $Q$ is encoded in the same manner. To perform pseudo-division, we shift the exponent of $Q$ by the difference between the leading degrees of $P$ and $Q$, aligning the terms to be eliminated. Quantum multiplication and subtraction are then applied to the matched coefficient values, canceling the leading term of $P$ and producing the point-value representation of the pseudo-remainder (Fig.~\ref{fig-Square}\textbf{d}). This construction reduces polynomial manipulation to arithmetic operations on pointwise values and avoids identifying and collecting like terms in a coefficient-based quantum representation. In the present implementation, the point values are precomputed classically and embedded into the controlled unitaries, whereas the arithmetic operations required for pseudo-division are executed on the quantum processor.

For the experimental implementation, the controlled unitaries $U_P$ and $U_Q$ in each step are compiled into elementary single- and two-qubit gates according to the available qubit connectivity and the required input-output relations. Auxiliary registers are introduced to store duplicated intermediate information, allowing part of the controlled operations to be executed in parallel and thereby reducing the circuit depth. For step 1, the resulting circuit is implemented on $21$ qubits and contains $77$ layers, including $150$ single-qubit gates and $89$ two-qubit gates.
In this step, the two polynomials are $P(\boldsymbol{x})=x_3x_5+(x_2x_4-x_1x_3)x_5^0$ and $Q(\boldsymbol{x})=x_5$, with $x_5$ as the leading variable to be eliminated. Since the exponent of $x_5$ takes only the values $0$ and $1$, the exponent register is encoded by a single qubit. The point register uses four qubits to encode the $N=2^4$ computational-basis states $\lvert x_1x_2x_3x_4\rangle$, where $x_i\in \{0,1 \}$. For each assignment, the controlled unitaries $U_P$ and $U_Q$ write the corresponding coefficient values of $P(\boldsymbol{x})$ and $Q(\boldsymbol{x})$ into two two-qubit value registers, whose ranges cover all coefficient values occurring in this step.
After encoding, pseudo-division is implemented through arithmetic multiplication and subtraction on the encoded coefficient values. This removes the highest-exponent term of $P(\boldsymbol{x})$, so that the coefficient of the exponent-one term in the resulting polynomial becomes zero in the output. The updated coefficients are read out by joint measurement of the exponent, point, and value registers (Supplementary Information Sec. 1).

The measurement outcomes of step 1 give the joint occupation probabilities of the seven measured qubits required to reconstruct the pseudo-remainder, shown as $2^7$ coloured squares in Fig.~\ref{fig-Square}\textbf{e}. The left and right groups correspond to the exponent qubit in states $|0\rangle$ and $|1\rangle$, respectively, while the x- and y-axes label the joint states of the point and value registers. The coefficient value represented by each two-qubit value state is indicated along the right-hand side. For each point index, the coefficient is identified from the value state with the largest probability, since the polynomial assigns a unique output to each input point. In the $|1\rangle$ sector, the values are zero for all sampled points, and the reconstructed coefficient is therefore zero, showing that the exponent-one term has been eliminated. In the $|0\rangle$ sector, the reconstructed values yield the output polynomial $R_1(\boldsymbol{x})=x_2x_4-x_1x_3$, consistent with the theoretical result of the pseudo-division step.
The output polynomial is then reordered with a new leading variable and processed with the next geometric equation. Repeating this procedure with the remaining hypothesis polynomials yields the final measurement results shown in Fig.~\ref{fig-Square}\textbf{f}. Additional results for steps 2 and 3 are provided in the Supplementary Information Sec.3. For each of the three exponent sectors and all three sampling points in the final step, the dominant probability is observed in the value state encoding zero, such that all reconstructed coefficients vanish and the final pseudo-remainder is $R_4(\boldsymbol{x})=0$. Together, the four QPD circuits realize a complete algebraic deduction of the desired conclusion that diagonals of a square are perpendicular.

We remark that a quantum implementation of Wu's method can, in principle, achieve a quadratic reduction in query complexity \cite{Sun2026Quantum}. In the present experiment, however, hardware imperfections limit the accessible circuit size and depth, and this advantage has not yet been demonstrated. Demonstrating the predicted scaling would require larger, deeper circuits supported by higher-fidelity quantum hardware and potentially fault-tolerant operations. We leave this for future study. 

\vspace{.5cm}
\noindent\textbf{\large{Quantum symbolic proof search}}

\begin{figure*}[htbp]
    \centering
    \includegraphics[width=\textwidth]{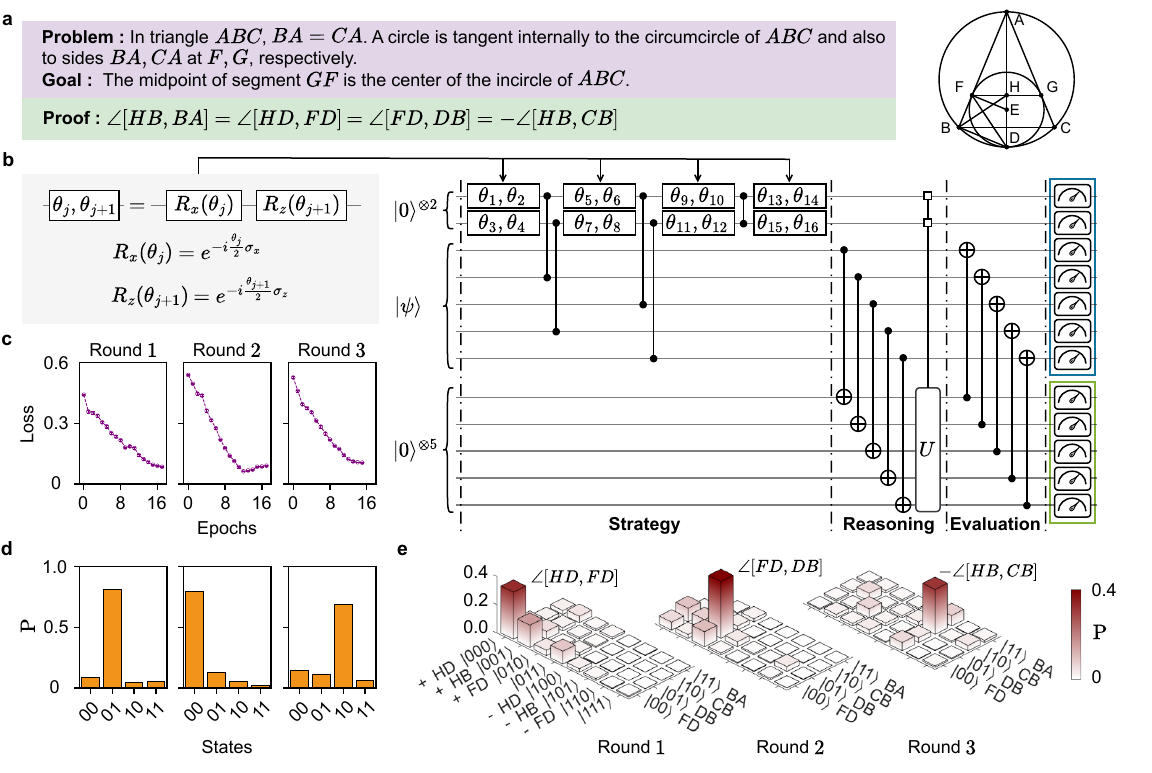}
    \caption{
        \textbf{Proving an IMO geometry theorem}.
        \textbf{a}, Problem 4 of the 1978 IMO, together with its geometric configuration and a proof expressed in the full-angle formalism.
        \textbf{b}, Quantum circuit for a single proof-search round. The circuit consists of a parameterized strategy circuit, a fixed reasoning circuit and an evaluation circuit. Two action qubits encode the candidate inference operations, five qubits encode the input proof state $|\psi\rangle$, and another five qubits encode the output proof state. The action register is initialized in $|0\rangle^{\otimes 2}$, and the output register in $|0\rangle^{\otimes 5}$. The input proof states are $|00111\rangle$, $|00000\rangle$ and $|01001\rangle$ in rounds 1, 2 and 3, respectively. The left gray box shows the parameterized single-qubit gate sequence on $\theta_j$ and $\theta_{j+1}$ used in the strategy circuit.
        \textbf{c}, Experimental training trajectories of the loss in the three proof-search rounds. Training is terminated when the standard deviation of the four most recent loss values falls below $0.005$. Error bars denote the standard error of the mean over 10 repeated runs, each with 1,000 shots. The loss is obtained from measurements of the registers enclosed by the blue box in \textbf{b}.
        \textbf{d}, Experimental probability distributions of the action-register states in the three rounds. The dominant outcomes select the three successive transformations used in the proof. Probabilities are obtained from 10 repeated measurements of the action register, each with 1,000 shots.
        \textbf{e}, Experimental probability distributions of the output full-angle states in the three rounds. The horizontal and vertical axes label the encoded sign and first line, and the second line, respectively, with the corresponding geometric meanings indicated along the axes. The dominant outputs are $\angle[HD,FD]$, $\angle[FD,DB]$ and $-\angle[HB,CB]$ in rounds 1, 2 and 3, respectively. Colours denote probabilities obtained from 10 repeated measurements of the registers enclosed by the green box in \textbf{b}, each with 1,000 shots.} \label{fig-IMO}
\end{figure*}

\noindent The above algebraic elimination follows a prescribed reduction order. We now introduce quantum symbolic proof search, in which the sequence of inference steps is determined as reasoning proceeds rather than prescribed in advance. At reasoning round $t$, the current stage of the derivation is encoded as a quantum proof state $\lvert S_t\rangle$, and formal inference rules define the admissible transitions between proof states. Based on $\lvert S_t\rangle$, the parameterized strategy circuit generates an action state over candidate inference operations. The action state controls a fixed reasoning circuit that coherently applies the corresponding rules and produces the candidate output state $\lvert S_{t+1}\rangle$. An evaluation circuit assesses the resulting transitions and defines the objective used to optimize the strategy circuit. The trainable strategy circuit enables flexible exploration of the proof space, whereas the fixed reasoning circuit restricts state transitions to those permitted by the formal inference rules. The quantum components operate coherently, preserving superpositions of inference operations and proof states throughout the reasoning process. Repeated reasoning steps generate a sequence of proof states that gives an explicit symbolic proof when the goal is reached (Fig.~\ref{fig-Overview}\textbf{d} and see Supplementary Information Sec. 2 for details).

We demonstrate this procedure on Problem 4 of the 1978 IMO (Fig.~\ref{fig-IMO}\textbf{a}). For an isosceles triangle $ABC$ with $AB=AC$, a circle is tangent to the sides $AB$ and $AC$ at $F$ and $G$ and internally tangent to the circumcircle. The goal is to show that the midpoint $H$ of $FG$ is the incenter of $ABC$. Let $D$ be the point of tangency of the two circles. A full-angle $\angle[u,v]$ is the directed angle from line $u$ to line $v$, taken modulo $\pi$. In the non-degenerate configuration shown, $H$ lies inside the triangle and, by symmetry, on the internal bisector at $A$. The remaining angle-bisector condition at $B$ is therefore expressed as $\angle[HB,BA]+\angle[HB,CB]=0$. From the relations derived classically from the hypotheses, we retain three equalities: $\angle[HD,FD]=\angle[FD,DB]$, $\angle[HB,BA]=\angle[HD,FD]$ and $\angle[FD,DB]=-\angle[HB,CB]$. The action states $\lvert00\rangle$, $\lvert01\rangle$ and $\lvert10\rangle$ encode these equalities, respectively. This leads to the proof sequence
\begin{align}\label{eq-IMO-proof}
\angle[HB,BA] \xrightarrow{\lvert01\rangle} \angle[HD,FD] & \xrightarrow{\lvert00\rangle} \angle[FD,DB] \\ \nonumber &\xrightarrow{\lvert10\rangle} -\angle[HB,CB].
\end{align}
The remaining basis state $\lvert11\rangle$ of the two-qubit action register is unused. To reduce the circuit overhead, the three selected relations are compiled directly into the reasoning circuit. Each action implements the corresponding equality as a directed transformation from its left-hand side to its right-hand side. The input and output proof states are stored in two five-qubit registers. In each register, the first qubit encodes the sign, and two pairs of qubits encode the two lines defining the full-angle  (Supplementary Information Sec. 2).

Each proof-search round comprises a parameterized strategy circuit, a fixed reasoning circuit and an evaluation circuit (Fig.~\ref{fig-IMO}\textbf{b}). In the strategy circuit, every action qubit is coupled to all four line-index qubits in the input proof-state register, thereby correlating action selection with the current proof state. The reasoning circuit first copies the input proof state to the output register and then conditionally performs the transformation specified by the action state. If the selected rule is not applicable to the input proof state, the output remains unchanged. The evaluation circuit compares the input and output proof states, favouring actions that produce a non-trivial state update while suppressing population in the unused state $\lvert11\rangle$. Its output defines the loss used to optimize the strategy circuit.

The proof search is initialized with $\angle[HB,BA]$. In each round, the strategy circuit is optimized for the current proof state, after which the action and output registers are measured. The dominant output full-angle is used as the input to the subsequent round. Three rounds are sufficient to complete the proof. The loss decreases throughout each optimization process (Fig.~\ref{fig-IMO}\textbf{c}), while the measured action distributions are dominated by $\lvert01\rangle$, $\lvert00\rangle$ and $\lvert10\rangle$ in the three successive rounds (Fig.~\ref{fig-IMO}\textbf{d}). The corresponding output distributions exhibit dominant peaks at $\angle[HD,FD]$, $\angle[FD,DB]$ and $-\angle[HB,CB]$ (Fig.~\ref{fig-IMO}\textbf{e}). These results reconstruct the sequence in Eq.~(\ref{eq-IMO-proof}) and establish the target identity $\angle[HB,BA]+\angle[HB,CB]=0$.

\vspace{.5cm}

\noindent\textbf{\large{Conclusion and outlook}}

\noindent We have experimentally demonstrated quantum automated theorem proving on a programmable superconducting processor through two complementary forms of formal reasoning. In the algebraic approach, polynomial deduction is mapped to quantum pseudo-division, and the conclusion is certified by the vanishing of the final pseudo-remainder. In the symbolic approach, proof states, inference actions and formal rules are encoded in quantum states and quantum circuits, allowing a proof path to be constructed through successive symbolic transformations. Using representative geometry problems, including a problem from the 1978 IMO, we implement both procedures on quantum hardware and obtain either an algebraic certificate or an explicit symbolic proof. These results show that structured mathematical reasoning can be formulated as an executable quantum process rather than merely verified through classical post-processing.

The present experiments operate on polynomial instances of limited size and a restricted set of symbolic relations, with the proof state measured and re-prepared between successive reasoning rounds. These restrictions arise from the available hardware resources rather than from the structure of the framework itself. Larger knowledge bases and richer rule sets can be incorporated by extending the state registers and compiling additional inference operations into the reasoning circuit. Passing the output proof state directly to subsequent reasoning steps would preserve superpositions of inference actions and proof paths throughout a multi-step derivation, allowing increasingly complex spaces of formally valid deductions to be processed within the same quantum architecture. By extending quantum computing beyond numerical tasks to verifiable symbolic reasoning, our work makes a first experimental step towards quantum automated theorem proving and, more broadly, quantum symbolic intelligence.
\vspace{.5cm}

\noindent\textbf{\large{Methods}}

\vspace{.3cm}
\noindent\textbf{Point-value representation and quantum pseudo-division}

\noindent For each elimination step, we partition the complete set of geometric variables as $\boldsymbol{x}=(\boldsymbol{z},y)$, where $y$ is the variable to be eliminated and $\boldsymbol{z}$ collects the remaining variables. The dividend $P$ and divisor $Q$ are expressed as univariate polynomials in $y$:

\begin{align}
P(\boldsymbol{z},y)
&=
\sum_{d=0}^{D_P}p_d(\boldsymbol{z})y^d,
&
Q(\boldsymbol{z},y)
&=
\sum_{d=0}^{D_Q}q_d(\boldsymbol{z})y^d,
\end{align}
where $D_P$ and $D_Q$ are their respective degrees in $y$, and $p_d(\boldsymbol{z})$ and $q_d(\boldsymbol{z})$ are coefficient polynomials in the remaining variables.

We represent these coefficient polynomials by reversible evaluation circuits \cite{Sun2026Quantum}. The circuit associated with $P$ implements
\begin{align}
U_{P,y}
\lvert d\rangle
\lvert\boldsymbol{z}\rangle
\lvert\boldsymbol{0}\rangle
=
\lvert d\rangle
\lvert\boldsymbol{z}\rangle
\lvert p_d(\boldsymbol{z})\rangle,
\end{align}
for the computational-basis inputs supported by the corresponding registers. The circuit $U_{Q,y}$ is defined analogously. Thus, the exponent register selects the coefficient polynomial associated with a given power of $y$, while the point register specifies the input of this polynomial. When the exponent and point registers are prepared in superposition, the corresponding coefficient values are evaluated coherently and written into the value register.

A pseudo-division reduction cancels the leading term of $P$ without dividing by the leading coefficient of $Q$. Defining $\Delta=D_P-D_Q$, one reduction produces
\begin{align}
R(\boldsymbol{z},y)
&=
q_{D_Q}(\boldsymbol{z})P(\boldsymbol{z},y)
-
p_{D_P}(\boldsymbol{z})y^{\Delta}Q(\boldsymbol{z},y)
\nonumber\\
&=
\sum_d r_d(\boldsymbol{z})y^d,
\end{align}
where
\begin{align}
r_d(\boldsymbol{z})
=
q_{D_Q}(\boldsymbol{z})p_d(\boldsymbol{z})
-
p_{D_P}(\boldsymbol{z})q_{d-\Delta}(\boldsymbol{z}),
\end{align}
with $q_k(\boldsymbol{z})=0$ for $k<0$ or $k>D_Q$. In particular, the leading coefficient satisfies $r_{D_P}(\boldsymbol{z})=0$, and the degree of the resulting polynomial in $y$ is reduced.

The quantum pseudo-division circuit evaluates this relation coherently. An exponent-shift operation first aligns the powers of $Q$ with those of $P$. The controlled circuits $U_{P,y}$ and $U_{Q,y}$ then write the coefficient values $p_d(\boldsymbol{z})$, $q_{D_Q}(\boldsymbol{z})$, $p_{D_P}(\boldsymbol{z})$ and $q_{d-\Delta}(\boldsymbol{z})$ into separate value registers. Reversible multiplication and subtraction operations subsequently implement

\begin{align}
\lvert d\rangle
\lvert\boldsymbol{z}\rangle
\lvert\boldsymbol{0}\rangle
\longmapsto
\lvert d\rangle
\lvert\boldsymbol{z}\rangle
\lvert r_d(\boldsymbol{z})\rangle,
\end{align}
with auxiliary registers retaining the intermediate values required by the arithmetic operations. If the degree of the resulting polynomial remains no smaller than $D_Q$, the reduction can be repeated until the degree of the pseudo-remainder in $y$ is lower than that of the divisor. For the polynomial system considered in this work, each variable-elimination step requires only one such reduction.

The polynomial-dependent evaluation circuits are constructed from the corresponding coefficient functions, whereas the exponent alignment and reversible arithmetic required to obtain the pseudo-remainder are executed on the quantum processor. The resulting pseudo-remainder retains the same point-value representation and can therefore be processed by subsequent polynomial operations within the same circuit framework. In the present experiment, each elimination step is implemented separately, with the point register prepared over a finite set of inputs sufficient to determine the relevant coefficient polynomials. The measured outputs are used to identify the intermediate pseudo-remainder and prepare the representation required for the next elimination step. The polynomial pairs and input values used in the four elimination steps, together with representative theoretical and compiled circuits, are provided in the Supplementary Information Sec. 1.

\vspace{.3cm}
\noindent\textbf{Quantum representation of full-angle formulas and inference actions}

\noindent A full-angle is specified by an ordered pair of lines \cite{Chou199612Automated}. In the experimental implementation, each proof state contains a signed full-angle and is represented as
\begin{align}
\lvert S\rangle
=
\lvert s\rangle
\lvert \ell_1\rangle
\lvert \ell_2\rangle,
\end{align}
where $\lvert s\rangle$ encodes the sign of the full-angle, and $\lvert \ell_1\rangle$ and $\lvert \ell_2\rangle$ encode its first and second lines, respectively. The sign is stored in qubit $Q^{S}_{1}$, whereas the two lines are stored in the two-qubit registers $Q^{S}_{2}Q^{S}_{3}$ and $Q^{S}_{4}Q^{S}_{5}$. A separate two-qubit register $Q^{A}_{1}Q^{A}_{2}$ represents the inference action. The computational-basis encoding used in the experiment is summarized in Table~\ref{table:encoding scheme}.

\begin{table}[htbp]
\begin{tabular}{ScScScSc}
\hline
\textbf{Name} & \textbf{Qubits} & \textbf{Value} & \textbf{Information}\\
\hline
\multirow{3}{*}{Action} & \multirow{3}{*}{$Q^{A}_{1}Q^{A}_{2}$} & 00 & $\angle[HD, FD]=\angle[FD, DB]$\\
 &  & 01 & $\angle[HB, BA]=\angle[HD, FD]$\\
 &  & 10 & $\angle[FD, DB]=-\angle[HB, CB]$\\
\hline
\multirow{9}{*}{State} & \multirow{2}{*}{$Q^{S}_{1}$} & 0 & $+$\\
 &  & 1 & $-$\\
\cline{2-4}
 & \multirow{3}{*}{$Q^{S}_{2}Q^{S}_{3}$} & 00 & $HD$\\
 &  & 01 & $HB$\\
 &  & 10 & $FD$\\
\cline{2-4}
 & \multirow{4}{*}{$Q^{S}_{4}Q^{S}_{5}$} & 00 & $FD$\\
 &  & 01 & $DB$\\
 &  & 10 & $CB$\\
 &  & 11 & $BA$\\
\hline
\end{tabular}
\caption{Basis encoding of the action and state registers. \label{table:encoding scheme}}
\end{table}

The three valid action states encode full-angle equalities obtained from the geometric hypotheses and inference rules. In the reasoning circuit, each equality is implemented as a directed transformation from its left-hand side to its right-hand side. The remaining action state $\lvert11\rangle$ is not assigned to an inference operation. For example, the initial proof state $\angle[HB,BA]$ is encoded as
\begin{align}
\lvert S_0\rangle
=
\lvert0\rangle
\lvert01\rangle
\lvert11\rangle
=
\lvert00111\rangle.
\end{align}
This problem-specific encoding retains only the full-angles and inference actions required for the implemented proof, thereby reducing the number of qubits and the circuit depth.

\vspace{.3cm}
\noindent\textbf{Quantum reasoning and evaluation circuits}

\noindent At reasoning round $t$, the strategy circuit prepares an action state $\lvert A_t\rangle$, which controls a fixed reasoning circuit acting on the input proof-state register $\lvert S_t\rangle$ and a five-qubit output register initialized in $\lvert\boldsymbol{0}\rangle$. On computational-basis states, the reasoning operation is defined by
\begin{align}
U_{\mathrm{R}}
\lvert a\rangle
\lvert S_t\rangle
\lvert\boldsymbol{0}\rangle
=
\lvert a\rangle
\lvert S_t\rangle
\lvert\widetilde{S}_{t+1}(a,S_t)\rangle,
\end{align}
where $\lvert\widetilde{S}_{t+1}(a,S_t)\rangle$ denotes the candidate proof state generated by action $a$. For each of the three valid actions, the corresponding full-angle equality in Table~\ref{table:encoding scheme} is applied from left to right when the input proof state matches its left-hand side. Otherwise, the output proof state remains unchanged. By linearity, the same circuit acts coherently on superpositions of proof states and inference actions.

The reasoning circuit uses two action qubits, five input-state qubits and five output-state qubits. Five CNOT gates first correlate the output register with the input register, such that each computational-basis component initially carries the same proof-state label. The action-conditioned unitary then updates the output register only when both the action and the input proof state match one of the compiled transformations. Its action can be summarized as
\begin{align}
\widetilde{S}_{t+1}(a,S_t)
=
\begin{cases}
f_a(S_t), & \text{if action $a$ is applicable to $S_t$},\\
S_t, & \text{otherwise},
\end{cases}
\end{align}
where $f_a$ denotes the directed full-angle transformation associated with action $a$. The transformations on the remaining basis states are chosen to complete the unitary operation. This construction separates action selection from formal inference: the trainable strategy circuit determines the action amplitudes, whereas the fixed reasoning circuit implements the corresponding rule-consistent state transitions.

The evaluation circuit assesses whether the candidate transition produces a new proof state and whether the selected action is valid. As shown in Fig.~\ref{fig-IMO}b, five CNOT gates are applied between the corresponding qubits of the input and output proof-state registers. These gates compute their bitwise difference in one of the registers, such that the measurement outcome $00000$ occurs if and only if the two encoded proof states are identical. We define the loss function as
\begin{align}
L
&=
P_1+P_2,
\end{align}
where $P_1$ and $P_2$ are probabilities estimated from repeated measurements of the corresponding registers. Specifically,
\begin{align}
P_1
&=
\Pr\!\left(S_t\oplus\widetilde{S}_{t+1}=00000\right),\\
P_2
&=
\Pr\!\left(A_t=11,S_t\oplus\widetilde{S}_{t+1} \ne 00000\right).
\end{align}
The probability $P_1$ quantifies the total probability weight of same outcomes in the joint measurement of the two proof-state registers. The probability $P_2$ quantifies the probability of obtaining the unused action outcome $11$ together with unequal outcomes for the two proof-state registers, thereby excluding events already counted in $P_1$. Minimizing $L$ therefore favours non-trivial proof-state transitions generated by the three valid actions. The resulting loss is used to optimize the parameters of the strategy circuit, while the candidate output obtained after optimization determines the proof state adopted for the subsequent reasoning round. Detailed action-conditioned transformations and gate-level constructions of the reasoning and evaluation circuits are provided in the Supplementary Information Sec.~2.

\vspace{.3cm}
\noindent\textbf{Optimization of the strategy circuit}

\noindent The strategy circuit contains 16 trainable parameters implemented as rotation angles of $R_x$ and $R_z$ gates, as shown in Fig.~3b. In each reasoning round, the circuit parameters are optimized using the Adam optimizer. The gradient of the loss with respect to each parameter is evaluated using the parameter-shift rule~\cite{Mitarai2018Quantum},
\begin{align}
\frac{\partial L}{\partial\theta_j}
=
\frac{
L(\theta_j+\pi/2)-L(\theta_j-\pi/2)
}{2}.
\end{align}

\vspace{.6cm}
\noindent\textbf{\large{}Data availability} \\
The data presented in the figures and that support the other findings of this study will be made publicly available for download on Zenodo upon publication.

\vspace{.6cm}
\noindent\textbf{\large{}Code availability} \\
The data analysis and numerical simulation codes will be made publicly available for download on Zenodo upon publication.

\vspace{.5cm}
\noindent\textbf{Acknowledgements} We are indebted to Vedran Dunjko for his invaluable comments and suggestions on the first version of the manuscript. We also thank Peter Zoller, Nobuyuki Yoshioka, Yasunobu Nakamura, Eric R. Anschuetz, and Haomu Yuan for discussions, and Jiachen Chen, Chuanyu Zhang, Yu Gao, Xuhao Zhu, and Feitong Jin for technical support. The device
was fabricated at the Micro-Nano Fabrication Center of Zhejiang University. We acknowledge the support from the National Natural Science Foundation
of China (grant nos. T2225008, 92565301, 12274368, 12404570, and 12404574), Quantum Science and Technology-National Science and Technology Major Project (grant nos. 2021ZD0300200 and 2021ZD0302203), Zhejiang Province Pioneer Plan (grant nos. 2025C01046, 2025C01019 and 2026C02A2004), and the Zhejiang Provincial Natural Science Foundation of China (grant no. LR24A040002). 
Z.-Z.S., Q.Y., S.J., W.L., Z.L., S.G., Y.M., and D.-L.D. acknowledge in addition support from the Tsinghua University Dushi Program and the Shanghai Qi Zhi Institute Innovation Program. Z.-Z. S. is additionally funded by China Postdoctoral Science Foundation (Certificated Number: 2025T180926).

\vspace{.5cm}
\noindent\textbf{Author contributions} 
{N.W. carried out the experiments and analyzed the experimental data under the supervision of C.S. and H.Wang;  Z.-Z.S., Q.Y., S.J., W.L., Y.M., S.G., Z.L., and D.-L.D. conducted the theoretical analysis; H.L. fabricated the device supervised by H.Wang; Z.-Z.S., N.W., C.S., H.Wang, and D.-L.D. co-wrote the manuscript; H.Wang, Q.G.,
C.S, P.Z.,
Z.C., Y.Z., A.Z., F.S., J.Z., Z.B., Z.Z., Han Wang, J.Y., J.S., G.L., Y.Wang, Y.-H.H., Y.-Y.H., J.H., 
S.Z., X.Z., Y.Wu, Z.-X.S., J.D., and H.D. contributed to experimental setup. All authors contributed to the discussions of the results.}

\vspace{.3cm}
\noindent\textbf{Competing interests}  All authors declare no competing interests.

\clearpage
\makeatother
\bibliography{Dengbib}

@article{Aaronson2014Need,
  author  = {Aaronson, Scott and Ambainis, Andris},
  title   = {The Need for Structure in Quantum Speedups},
  journal = {Theory of Computing},
  year    = {2014},
  volume  = {10},
  number  = {6},
  pages   = {133--166},
  doi     = {10.4086/toc.2014.v010a006},
}

@inproceedings{Alemi2016DeepMath,
  author    = {Alemi, Alexander A. and Chollet, Fran\c{c}ois and Een, Niklas and Irving, Geoffrey and Szegedy, Christian and Urban, Josef},
  title     = {{DeepMath}---Deep Sequence Models for Premise Selection},
  booktitle = {Advances in Neural Information Processing Systems 29},
  year      = {2016},
  pages     = {2235--2243},
  publisher = {Curran Associates, Inc.},
}

@article{Barenco199511Elementary,
  author  = {Barenco, Adriano and Bennett, Charles H. and Cleve, Richard and DiVincenzo, David P. and Margolus, Norman and Shor, Peter and Sleator, Tycho and Smolin, John A. and Weinfurter, Harald},
  title   = {Elementary Gates for Quantum Computation},
  journal = {Physical Review A},
  year    = {1995},
  volume  = {52},
  number  = {5},
  pages   = {3457--3467},
  doi     = {10.1103/PhysRevA.52.3457},
}

@article{Chou198809introduction,
  author  = {Chou, Shang-Ching},
  title   = {An Introduction to {Wu}'s Method for Mechanical Theorem Proving in Geometry},
  journal = {Journal of Automated Reasoning},
  year    = {1988},
  volume  = {4},
  number  = {3},
  pages   = {237--267},
  doi     = {10.1007/BF00244942},
}

@article{Chou199612Automated,
  author  = {Chou, Shang-Ching and Gao, Xiao-Shan and Zhang, Jing-Zhong},
  title   = {Automated Generation of Readable Proofs with Geometric Invariants: {II}. Theorem Proving with Full-Angles},
  journal = {Journal of Automated Reasoning},
  year    = {1996},
  volume  = {17},
  number  = {3},
  pages   = {349--370},
  doi     = {10.1007/BF00283134},
}

@misc{Coppersmith2002approximate,
  author        = {Coppersmith, D.},
  title         = {An Approximate {Fourier} Transform Useful in Quantum Factoring},
  year          = {2002},
  eprint        = {quant-ph/0201067},
  archiveprefix = {arXiv},
  primaryclass  = {quant-ph},
  doi           = {10.48550/arXiv.quant-ph/0201067},
  url           = {https://arxiv.org/abs/quant-ph/0201067},
  note          = {{IBM} Research Report RC 19642},
}

@article{Deutsch199212Rapid,
  author  = {Deutsch, David and Jozsa, Richard},
  title   = {Rapid Solution of Problems by Quantum Computation},
  journal = {Proceedings of the Royal Society of London. Series A: Mathematical and Physical Sciences},
  year    = {1992},
  volume  = {439},
  number  = {1907},
  pages   = {553--558},
  doi     = {10.1098/rspa.1992.0167},
}

@book{Fitting2012First,
  author    = {Fitting, Melvin},
  title     = {First-Order Logic and Automated Theorem Proving},
  edition   = {2},
  series    = {Graduate Texts in Computer Science},
  year      = {1996},
  publisher = {Springer-Verlag},
  address   = {New York, NY},
  isbn      = {978-0-387-94593-4},
  doi       = {10.1007/978-1-4612-2360-3},
}

@article{Gowers2025Conjecture,
  author  = {Gowers, William Timothy and Green, Ben and Manners, Freddie and Tao, Terence},
  title   = {On a Conjecture of {Marton}},
  journal = {Annals of Mathematics},
  year    = {2025},
  volume  = {201},
  number  = {2},
  pages   = {515--549},
  doi     = {10.4007/annals.2025.201.2.5},
}

@inproceedings{Grover1996fast,
  author    = {Grover, Lov K.},
  title     = {A Fast Quantum Mechanical Algorithm for Database Search},
  booktitle = {Proceedings of the Twenty-Eighth Annual ACM Symposium on Theory of Computing},
  year      = {1996},
  pages     = {212--219},
  publisher = {Association for Computing Machinery},
  address   = {New York, NY},
  isbn      = {978-0-89791-785-8},
  doi       = {10.1145/237814.237866},
}

@incollection{Hasan2015Formal,
  author    = {Hasan, Osman and Tahar, Sofi\`ene},
  title     = {Formal Verification Methods},
  booktitle = {Encyclopedia of Information Science and Technology, Third Edition},
  editor    = {Khosrow-Pour, Mehdi},
  year      = {2015},
  pages     = {7162--7170},
  publisher = {IGI Global},
  address   = {Hershey, PA},
  isbn      = {978-1-4666-5888-2},
  doi       = {10.4018/978-1-4666-5888-2.ch705},
}

@inproceedings{Kabanets2003Derandomizing,
  author    = {Kabanets, Valentine and Impagliazzo, Russell},
  title     = {Derandomizing Polynomial Identity Tests Means Proving Circuit Lower Bounds},
  booktitle = {Proceedings of the Thirty-Fifth Annual ACM Symposium on Theory of Computing},
  year      = {2003},
  pages     = {355--364},
  publisher = {Association for Computing Machinery},
  address   = {New York, NY},
  doi       = {10.1145/780542.780595},
}

@inproceedings{Kaliszyk2018Reinforcement,
  author    = {Kaliszyk, Cezary and Urban, Josef and Michalewski, Henryk and Ol\v{s}\'ak, Miroslav},
  title     = {Reinforcement Learning of Theorem Proving},
  booktitle = {Advances in Neural Information Processing Systems 31},
  year      = {2018},
  pages     = {8822--8833},
  publisher = {Curran Associates, Inc.},
}

@book{Loveland2016Automated,
  author    = {Loveland, Donald W.},
  title     = {Automated Theorem Proving: A Logical Basis},
  series    = {Fundamental Studies in Computer Science},
  volume    = {6},
  edition   = {1},
  year      = {1978},
  publisher = {North-Holland Publishing Company},
  address   = {Amsterdam},
  isbn      = {978-0-7204-0499-9},
}

@article{Nevins1975Plane,
  author  = {Nevins, Arthur J.},
  title   = {Plane Geometry Theorem Proving Using Forward Chaining},
  journal = {Artificial Intelligence},
  year    = {1975},
  volume  = {6},
  number  = {1},
  pages   = {1--23},
  doi     = {10.1016/0004-3702(75)90013-2},
}

@book{Nielsen201206Quantum,
  author    = {Nielsen, Michael A. and Chuang, Isaac L.},
  title     = {Quantum Computation and Quantum Information},
  edition   = {10th Anniversary Edition},
  year      = {2010},
  publisher = {Cambridge University Press},
  address   = {Cambridge},
  isbn      = {978-1-107-00217-3},
  doi       = {10.1017/CBO9780511976667},
}

@article{RuizPerez201704Quantum,
  author  = {Ruiz-P\'erez, Lidia and Garc\'ia-Escart\'in, Juan Carlos},
  title   = {Quantum Arithmetic with the Quantum {Fourier} Transform},
  journal = {Quantum Information Processing},
  year    = {2017},
  volume  = {16},
  number  = {6},
  pages   = {152},
  doi     = {10.1007/s11128-017-1603-1},
}

@inproceedings{Shor1994Algorithms,
  author    = {Shor, Peter W.},
  title     = {Algorithms for Quantum Computation: Discrete Logarithms and Factoring},
  booktitle = {Proceedings of the 35th Annual Symposium on Foundations of Computer Science},
  year      = {1994},
  pages     = {124--134},
  publisher = {IEEE Computer Society Press},
  doi       = {10.1109/SFCS.1994.365700},
}

@article{Simon199710Power,
  author  = {Simon, Daniel R.},
  title   = {On the Power of Quantum Computation},
  journal = {SIAM Journal on Computing},
  year    = {1997},
  volume  = {26},
  number  = {5},
  pages   = {1474--1483},
  doi     = {10.1137/S0097539796298637},
}

@article{Stahlke201408Quantum,
  author  = {Stahlke, Dan},
  title   = {Quantum Interference as a Resource for Quantum Speedup},
  journal = {Physical Review A},
  year    = {2014},
  volume  = {90},
  number  = {2},
  pages   = {022302},
  doi     = {10.1103/PhysRevA.90.022302},
}

@misc{Sun2026Quantum,
  author        = {Sun, Zheng-Zhi and Ye, Qi and Deng, Dong-Ling},
  title         = {Quantum Automated Theorem Proving},
  year          = {2026},
  eprint        = {2601.07953},
  archiveprefix = {arXiv},
  primaryclass  = {quant-ph},
  doi           = {10.48550/arXiv.2601.07953},
  url           = {https://arxiv.org/abs/2601.07953},
}

@article{Wu1978Decision,
  author  = {Wu, Wen-Tsun},
  title   = {On the Decision Problem and the Mechanization of Theorem-Proving in Elementary Geometry},
  journal = {Scientia Sinica},
  year    = {1978},
  volume  = {21},
  number  = {2},
  pages   = {159--172},
}

@article{Yoder201411Fixed,
  author  = {Yoder, Theodore J. and Low, Guang Hao and Chuang, Isaac L.},
  title   = {Fixed-Point Quantum Search with an Optimal Number of Queries},
  journal = {Physical Review Letters},
  year    = {2014},
  volume  = {113},
  number  = {21},
  pages   = {210501},
  doi     = {10.1103/PhysRevLett.113.210501},
}

@article{Acharya202302Suppressing,
  author = {{Google Quantum AI}},
  title = {Suppressing quantum errors by scaling a surface code logical qubit},
  journal = {Nature},
  year = {2023},
  volume = {614},
  number = {7949},
  pages = {676--681},
  doi = {10.1038/s41586-022-05434-1},
}

@article{Acharya2025,
  author = {{Google Quantum AI and Collaborators}},
  title = {Quantum error correction below the surface code threshold},
  journal = {Nature},
  year = {2025},
  volume = {638},
  number = {8052},
  pages = {920--926},
  doi = {10.1038/s41586-024-08449-y},
}

@article{Arute2019Quantum,
  author = {Arute, Frank and Arya, Kunal and Babbush, Ryan and Bacon, Dave and Bardin, Joseph C. and Barends, Rami and Biswas, Rupak and Boixo, Sergio and Brandao, Fernando G. S. L. and Buell, David A. and others},
  title = {Quantum supremacy using a programmable superconducting processor},
  journal = {Nature},
  year = {2019},
  volume = {574},
  number = {7779},
  pages = {505--510},
  doi = {10.1038/s41586-019-1666-5},
}

@article{Bluvstein2024Logical,
  author = {Bluvstein, Dolev and Evered, Simon J. and Geim, Alexandra A. and Li, Sophie H. and Zhou, Hengyun and Manovitz, Tom and Ebadi, Sepehr and Cain, Madelyn and Kalinowski, Marcin and Hangleiter, Dominik and others},
  title = {Logical quantum processor based on reconfigurable atom arrays},
  journal = {Nature},
  year = {2024},
  volume = {626},
  number = {7997},
  pages = {58--65},
  doi = {10.1038/s41586-023-06927-3},
}

@article{Chen2016Measuring,
  author = {Chen, Zijun and Kelly, Julian and Quintana, Chris and Barends, R. and Campbell, B. and Chen, Yu and Chiaro, B. and Dunsworth, A. and Fowler, A. G. and Lucero, E. and others},
  title = {Measuring and Suppressing Quantum State Leakage in a Superconducting Qubit},
  journal = {Physical Review Letters},
  year = {2016},
  volume = {116},
  number = {2},
  pages = {020501},
  doi = {10.1103/PhysRevLett.116.020501},
}

@article{Ebadi202107Quantum,
  author = {Ebadi, Sepehr and Wang, Tout T. and Levine, Harry and Keesling, Alexander and Semeghini, Giulia and Omran, Ahmed and Bluvstein, Dolev and Samajdar, Rhine and Pichler, Hannes and Ho, Wen Wei and others},
  title = {Quantum phases of matter on a 256-atom programmable quantum simulator},
  journal = {Nature},
  year = {2021},
  volume = {595},
  number = {7866},
  pages = {227--232},
  doi = {10.1038/s41586-021-03582-4},
}

@article{Egan202110Fault,
  author = {Egan, Laird and Debroy, Dripto M. and Noel, Crystal and Risinger, Andrew and Zhu, Daiwei and Biswas, Debopriyo and Newman, Michael and Li, Muyuan and Brown, Kenneth R. and Cetina, Marko and others},
  title = {Fault-tolerant control of an error-corrected qubit},
  journal = {Nature},
  year = {2021},
  volume = {598},
  number = {7880},
  pages = {281--286},
  doi = {10.1038/s41586-021-03928-y},
}

@article{Google2025Observation,
  author = {{Google Quantum AI and Collaborators}},
  title = {Observation of constructive interference at the edge of quantum ergodicity},
  journal = {Nature},
  year = {2025},
  volume = {646},
  number = {8086},
  pages = {825--830},
  doi = {10.1038/s41586-025-09526-6},
}

@article{Gupta2024Encoding,
  author = {Gupta, Riddhi S. and Sundaresan, Neereja and Alexander, Thomas and Wood, Christopher J. and Merkel, Seth T. and Healy, Michael B. and Hillenbrand, Marius and Jochym-O'Connor, Tomas and Wootton, James R. and Yoder, Theodore J. and others},
  title = {Encoding a magic state with beyond break-even fidelity},
  journal = {Nature},
  year = {2024},
  volume = {625},
  number = {7994},
  pages = {259--263},
  doi = {10.1038/s41586-023-06846-3},
}

@article{Jin2025Topological,
  author = {Jin, Feitong and Jiang, Si and Zhu, Xuhao and Bao, Zehang and Shen, Fanhao and Wang, Ke and Zhu, Zitian and Xu, Shibo and Song, Zixuan and Chen, Jiachen and others},
  title = {Topological prethermal strong zero modes on superconducting processors},
  journal = {Nature},
  year = {2025},
  volume = {645},
  number = {8081},
  pages = {626--632},
  doi = {10.1038/s41586-025-09476-z},
}

@article{Kim202306Evidence,
  author = {Kim, Youngseok and Eddins, Andrew and Anand, Sajant and Wei, Ken Xuan and van den Berg, Ewout and Rosenblatt, Sami and Nayfeh, Hasan and Wu, Yantao and Zaletel, Michael and Temme, Kristan and others},
  title = {Evidence for the utility of quantum computing before fault tolerance},
  journal = {Nature},
  year = {2023},
  volume = {618},
  number = {7965},
  pages = {500--505},
  doi = {10.1038/s41586-023-06096-3},
}

@article{King2025Beyond,
  author = {King, Andrew D. and Nocera, Alberto and Rams, Marek M. and Dziarmaga, Jacek and Wiersema, Roeland and Bernoudy, William and Raymond, Jack and Kaushal, Nitin and Heinsdorf, Niclas and Harris, Richard and others},
  title = {Beyond-classical computation in quantum simulation},
  journal = {Science},
  year = {2025},
  volume = {388},
  number = {6743},
  pages = {199--204},
  doi = {10.1126/science.ado6285},
}

@article{Krinner202205Realizing,
  author = {Krinner, Sebastian and Lacroix, Nathan and Remm, Ants and Di Paolo, Agustin and Genois, Elie and Leroux, Catherine and Hellings, Christoph and Lazar, Stefania and Swiadek, Francois and Herrmann, Johannes and others},
  title = {Realizing repeated quantum error correction in a distance-three surface code},
  journal = {Nature},
  year = {2022},
  volume = {605},
  number = {7911},
  pages = {669--674},
  doi = {10.1038/s41586-022-04566-8},
}

@article{Lacroix2025Scaling,
  author = {Lacroix, N. and Bourassa, A. and Heras, F. J. H. and Zhang, L. M. and Bausch, J. and Senior, A. W. and Edlich, T. and Shutty, N. and Sivak, V. and Bengtsson, A. and others},
  title = {Scaling and logic in the colour code on a superconducting quantum processor},
  journal = {Nature},
  year = {2025},
  volume = {645},
  number = {8081},
  pages = {614--619},
  doi = {10.1038/s41586-025-09061-4},
}

@article{Madsen2022Quantum,
  author = {Madsen, Lars S. and Laudenbach, Fabian and Askarani, Mohsen Falamarzi and Rortais, Fabien and Vincent, Trevor and Bulmer, Jacob F. F. and Miatto, Filippo M. and Neuhaus, Leonhard and Helt, Lukas G. and Collins, Matthew J. and others},
  title = {Quantum computational advantage with a programmable photonic processor},
  journal = {Nature},
  year = {2022},
  volume = {606},
  number = {7912},
  pages = {75--81},
  doi = {10.1038/s41586-022-04725-x},
}

@article{Moses202312Race,
  author = {Moses, S. A. and Baldwin, C. H. and Allman, M. S. and Ancona, R. and Ascarrunz, L. and Barnes, C. and Bartolotta, J. and Bjork, B. and Blanchard, P. and Bohn, M. and others},
  title = {A Race-Track Trapped-Ion Quantum Processor},
  journal = {Physical Review X},
  year = {2023},
  volume = {13},
  number = {4},
  pages = {041052},
  doi = {10.1103/PhysRevX.13.041052},
}

@article{Ni202303Beating,
  author = {Ni, Zhongchu and Li, Sai and Deng, Xiaowei and Cai, Yanyan and Zhang, Libo and Wang, Weiting and Yang, Zhen-Biao and Yu, Haifeng and Yan, Fei and Liu, Song and others},
  title = {Beating the break-even point with a discrete-variable-encoded logical qubit},
  journal = {Nature},
  year = {2023},
  volume = {616},
  number = {7955},
  pages = {56--60},
  doi = {10.1038/s41586-023-05784-4},
}

@article{Sivak202303Real,
  author = {Sivak, V. V. and Eickbusch, A. and Royer, B. and Singh, S. and Tsioutsios, I. and Ganjam, S. and Miano, A. and Brock, B. L. and Ding, A. Z. and Frunzio, L. and others},
  title = {Real-time quantum error correction beyond break-even},
  journal = {Nature},
  year = {2023},
  volume = {616},
  number = {7955},
  pages = {50--55},
  doi = {10.1038/s41586-023-05782-6},
}

@article{Wang2026Demonstration,
  author = {Wang, Ke and Lu, Zhide and Zhang, Chuanyu and Liu, Gongyu and Chen, Jiachen and Wang, Yanzhe and Wu, Yaozu and Xu, Shibo and Zhu, Xuhao and Jin, Feitong and others},
  title = {Demonstration of low-overhead quantum error correction codes},
  journal = {Nature Physics},
  year = {2026},
  volume = {22},
  number = {2},
  pages = {308--314},
  doi = {10.1038/s41567-025-03157-4},
}

@article{Wu2021Strong,
  author = {Wu, Yulin and Bao, Wan-Su and Cao, Sirui and Chen, Fusheng and Chen, Ming-Cheng and Chen, Xiawei and Chung, Tung-Hsun and Deng, Hui and Du, Yajie and Fan, Daojin and others},
  title = {Strong Quantum Computational Advantage Using a Superconducting Quantum Processor},
  journal = {Physical Review Letters},
  year = {2021},
  volume = {127},
  number = {18},
  pages = {180501},
  doi = {10.1103/PhysRevLett.127.180501},
}

@article{Xu202407Non,
  author = {Xu, Shibo and Sun, Zheng-Zhi and Wang, Ke and Li, Hekang and Zhu, Zitian and Dong, Hang and Deng, Jinfeng and Zhang, Xu and Chen, Jiachen and Wu, Yaozu and others},
  title = {{Non-Abelian} braiding of {Fibonacci} anyons with a superconducting processor},
  journal = {Nature Physics},
  year = {2024},
  volume = {20},
  number = {9},
  pages = {1469--1475},
  doi = {10.1038/s41567-024-02529-6},
}

@article{Zhong202012Quantum,
  author = {Zhong, Han-Sen and Wang, Hui and Deng, Yu-Hao and Chen, Ming-Cheng and Peng, Li-Chao and Luo, Yi-Han and Qin, Jian and Wu, Dian and Ding, Xing and Hu, Yi and others},
  title = {Quantum computational advantage using photons},
  journal = {Science},
  year = {2020},
  volume = {370},
  number = {6523},
  pages = {1460--1463},
  doi = {10.1126/science.abe8770},
}

@inproceedings{Li2019Tackling,
  author = {Li, Gushu and Ding, Yufei and Xie, Yuan},
  title = {Tackling the Qubit Mapping Problem for {NISQ}-Era Quantum Devices},
  booktitle = {Proceedings of the Twenty-Fourth International Conference on Architectural Support for Programming Languages and Operating Systems},
  year = {2019},
  pages = {1001--1014},
  publisher = {ACM},
  doi = {10.1145/3297858.3304023},
}

@article{Maslov2016Advantages,
  author = {Maslov, Dmitri},
  title = {Advantages of using relative-phase {Toffoli} gates with an application to multiple control {Toffoli} optimization},
  journal = {Physical Review A},
  year = {2016},
  volume = {93},
  number = {2},
  pages = {022311},
  doi = {10.1103/PhysRevA.93.022311},
}

@article{Biamonte2017Quantum,
  author  = {Biamonte, Jacob and Wittek, Peter and Pancotti, Nicola and Rebentrost, Patrick and Wiebe, Nathan and Lloyd, Seth},
  title   = {Quantum machine learning},
  journal = {Nature},
  year    = {2017},
  volume  = {549},
  number  = {7671},
  pages   = {195--202},
  doi     = {10.1038/nature23474},
}

@article{Cerezo2022Challenges,
  author  = {Cerezo, M. and Verdon, Guillaume and Huang, Hsin-Yuan and Cincio, Lukasz and Coles, Patrick J.},
  title   = {Challenges and opportunities in quantum machine learning},
  journal = {Nature Computational Science},
  year    = {2022},
  volume  = {2},
  number  = {9},
  pages   = {567--576},
  doi     = {10.1038/s43588-022-00311-3},
}

@article{Collins2024Evaluating,
  author  = {Collins, Katherine M. and Jiang, Albert Q. and Frieder, Simon and Wong, Lionel and Zilka, Miri and Bhatt, Umang and Lukasiewicz, Thomas and Wu, Yuhuai and Tenenbaum, Joshua B. and Hart, William and others},
  title   = {Evaluating language models for mathematics through interactions},
  journal = {Proceedings of the National Academy of Sciences},
  year    = {2024},
  volume  = {121},
  number  = {24},
  pages   = {e2318124121},
  doi     = {10.1073/pnas.2318124121},
}

@article{DasSarma2019Machine,
  author  = {Das Sarma, Sankar and Deng, Dong-Ling and Duan, Lu-Ming},
  title   = {Machine learning meets quantum physics},
  journal = {Physics Today},
  year    = {2019},
  volume  = {72},
  number  = {3},
  pages   = {48--54},
  doi     = {10.1063/PT.3.4164},
}

@article{Dunjko2016Quantum,
  author  = {Dunjko, Vedran and Taylor, Jacob M. and Briegel, Hans J.},
  title   = {Quantum-enhanced machine learning},
  journal = {Physical Review Letters},
  year    = {2016},
  volume  = {117},
  number  = {13},
  pages   = {130501},
  doi     = {10.1103/PhysRevLett.117.130501},
}

@article{Dunjko2018Machine,
  author  = {Dunjko, Vedran and Briegel, Hans J.},
  title   = {Machine learning \& artificial intelligence in the quantum domain: a review of recent progress},
  journal = {Reports on Progress in Physics},
  year    = {2018},
  volume  = {81},
  number  = {7},
  pages   = {074001},
  doi     = {10.1088/1361-6633/aab406},
}

@article{Gao2018Quantum,
  author  = {Gao, Xun and Zhang, Zhengyu and Duan, Lu-Ming},
  title   = {A quantum machine learning algorithm based on generative models},
  journal = {Science Advances},
  year    = {2018},
  volume  = {4},
  number  = {12},
  pages   = {eaat9004},
  doi     = {10.1126/sciadv.aat9004},
}

@article{Guo2025DeepSeek,
  author  = {Guo, Daya and Yang, Dejian and Zhang, Haowei and Song, Junxiao and Wang, Peiyi and Zhu, Qihao and Xu, Runxin and Zhang, Ruoyu and Ma, Shirong and Bi, Xiao and others},
  title   = {{DeepSeek-R1} incentivizes reasoning in {LLMs} through reinforcement learning},
  journal = {Nature},
  year    = {2025},
  volume  = {645},
  number  = {8081},
  pages   = {633--638},
  doi     = {10.1038/s41586-025-09422-z},
}

@article{Harrow2009Quantum,
  author  = {Harrow, Aram W. and Hassidim, Avinatan and Lloyd, Seth},
  title   = {Quantum algorithm for linear systems of equations},
  journal = {Physical Review Letters},
  year    = {2009},
  volume  = {103},
  number  = {15},
  pages   = {150502},
  doi     = {10.1103/PhysRevLett.103.150502},
}

@article{Havlicek2019Supervised,
  author  = {Havl{\'i}{\v c}ek, Vojt{\v e}ch and C{\'o}rcoles, Antonio D. and Temme, Kristan and Harrow, Aram W. and Kandala, Abhinav and Chow, Jerry M. and Gambetta, Jay M.},
  title   = {Supervised learning with quantum-enhanced feature spaces},
  journal = {Nature},
  year    = {2019},
  volume  = {567},
  number  = {7747},
  pages   = {209--212},
  doi     = {10.1038/s41586-019-0980-2},
}

@article{Hu2019Quantum,
  author  = {Hu, Ling and Wu, Shu-Hao and Cai, Weizhou and Ma, Yuwei and Mu, Xianghao and Xu, Yuan and Wang, Haiyan and Song, Yipu and Deng, Dong-Ling and Zou, Chang-Ling and others},
  title   = {Quantum generative adversarial learning in a superconducting quantum circuit},
  journal = {Science Advances},
  year    = {2019},
  volume  = {5},
  number  = {1},
  pages   = {eaav2761},
  doi     = {10.1126/sciadv.aav2761},
}

@article{Huang2022Quantum,
  author  = {Huang, Hsin-Yuan and Broughton, Michael and Cotler, Jordan and Chen, Sitan and Li, Jerry and Mohseni, Masoud and Neven, Hartmut and Babbush, Ryan and Kueng, Richard and Preskill, John and others},
  title   = {Quantum advantage in learning from experiments},
  journal = {Science},
  year    = {2022},
  volume  = {376},
  number  = {6598},
  pages   = {1182--1186},
  doi     = {10.1126/science.abn7293},
}

@article{Hubert2026Olympiad,
  author  = {Hubert, Thomas and Mehta, Rishi and Sartran, Laurent and Horv{\'a}th, Mikl{\'o}s Z. and {\v Z}u{\v z}i{\'c}, Goran and Wieser, Eric and Huang, Aja and Schrittwieser, Julian and Schroecker, Yannick and Masoom, Hussain and others},
  title   = {Olympiad-level formal mathematical reasoning with reinforcement learning},
  journal = {Nature},
  year    = {2026},
  volume  = {651},
  number  = {8106},
  pages   = {607--613},
  doi     = {10.1038/s41586-025-09833-y},
}

@article{Jerbi2023Quantum,
  author  = {Jerbi, Sofiene and Fiderer, Lukas J. and Poulsen Nautrup, Hendrik and K{\"u}bler, Jonas M. and Briegel, Hans J. and Dunjko, Vedran},
  title   = {Quantum machine learning beyond kernel methods},
  journal = {Nature Communications},
  year    = {2023},
  volume  = {14},
  number  = {1},
  pages   = {517},
  doi     = {10.1038/s41467-023-36159-y},
}

@article{Li2025Pitfalls,
  author  = {Li, Weikang and Ma, Yixuan and Deng, Dong-Ling},
  title   = {Pitfalls and prospects of quantum machine learning},
  journal = {Nature Computational Science},
  year    = {2025},
  volume  = {5},
  number  = {12},
  pages   = {1095--1097},
  doi     = {10.1038/s43588-025-00914-6},
}

@article{Liu2021Rigorous,
  author  = {Liu, Yunchao and Arunachalam, Srinivasan and Temme, Kristan},
  title   = {A rigorous and robust quantum speed-up in supervised machine learning},
  journal = {Nature Physics},
  year    = {2021},
  volume  = {17},
  number  = {9},
  pages   = {1013--1017},
  doi     = {10.1038/s41567-021-01287-z},
}

@article{Lloyd2014Quantum,
  author  = {Lloyd, Seth and Mohseni, Masoud and Rebentrost, Patrick},
  title   = {Quantum principal component analysis},
  journal = {Nature Physics},
  year    = {2014},
  volume  = {10},
  number  = {9},
  pages   = {631--633},
  doi     = {10.1038/nphys3029},
}

@article{Ren2022Experimental,
  author  = {Ren, Wenhui and Li, Weikang and Xu, Shibo and Wang, Ke and Jiang, Wenjie and Jin, Feitong and Zhu, Xuhao and Chen, Jiachen and Song, Zixuan and Zhang, Pengfei and others},
  title   = {Experimental quantum adversarial learning with programmable superconducting qubits},
  journal = {Nature Computational Science},
  year    = {2022},
  volume  = {2},
  number  = {11},
  pages   = {711--717},
  doi     = {10.1038/s43588-022-00351-9},
}

@misc{Ren2025DeepSeek,
  author        = {Ren, Z. Z. and Shao, Zhihong and Song, Junxiao and Xin, Huajian and Wang, Haocheng and Zhao, Wanjia and Zhang, Liyue and Fu, Zhe and Zhu, Qihao and Yang, Dejian and others},
  title         = {{DeepSeek-Prover-V2}: Advancing formal mathematical reasoning via reinforcement learning for subgoal decomposition},
  year          = {2025},
  eprint        = {2504.21801},
  archivePrefix = {arXiv},
  primaryClass  = {cs.CL},
  doi           = {10.48550/arXiv.2504.21801},
  url           = {https://arxiv.org/abs/2504.21801}
}

@article{Trinh2024Solving,
  author  = {Trinh, Trieu H. and Wu, Yuhuai and Le, Quoc V. and He, He and Luong, Thang},
  title   = {Solving olympiad geometry without human demonstrations},
  journal = {Nature},
  year    = {2024},
  volume  = {625},
  number  = {7995},
  pages   = {476--482},
  doi     = {10.1038/s41586-023-06747-5},
}

@article{Zhang2026Proposing,
  author  = {Zhang, Chi and Song, Jiajun and Li, Siyu and Liang, Yitao and Ma, Yuxi and Wang, Wei and Zhu, Yixin and Zhu, Song-Chun},
  title   = {Proposing and solving olympiad geometry with guided tree search},
  journal = {Nature Machine Intelligence},
  year    = {2026},
  volume  = {8},
  number  = {1},
  pages   = {84--95},
  doi     = {10.1038/s42256-025-01164-x},
}

@article{Mitarai2018Quantum,
  title = {Quantum Circuit Learning},
  author = {Mitarai, K. and Negoro, M. and Kitagawa, M. and Fujii, K.},
  journal = {Physical Review A},
  volume = {98},
  issue = {3},
  pages = {032309},
  year = 2018,
  month = {Sep},
  publisher = {American Physical Society},
  doi = {10.1103/PhysRevA.98.032309}
}

\end{document}


\beginsupplement
\title{Supplementary Information: Proving olympiad geometry theorems on a superconducting quantum processor}	
\maketitle
\tableofcontents

\onecolumngrid

\section{Quantum algebraic theorem proving}

\begin{figure}[htb]
    \includegraphics[width=0.6\linewidth]
    {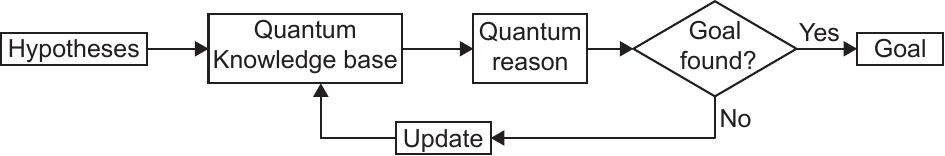}
    \caption{\label{fig-forward-QATP}Schematic reasoning cycle in QATP.
    The hypotheses are encoded in a quantum knowledge base and processed
    by quantum reasoning. The ``Goal found?'' test checks whether the
    resulting derivation establishes the target statement. If so, the
    procedure terminates successfully. Otherwise, intermediate results
    are used to update the knowledge base for the next round. The cycle
    continues until the target is established or a prescribed stopping
    condition is met.}
\end{figure}

We begin with Wu's method as an algebraic realization of quantum
automated theorem proving (QATP)~\cite{Sun2026Quantum}.
Fig.~\ref{fig-forward-QATP} summarizes the underlying reasoning cycle.
Given a set of hypotheses and a target statement, the hypotheses are
encoded in a quantum knowledge base. Quantum reasoning then derives new
formulas or transforms the symbolic expressions used in the proof. The
result is checked against the criterion for establishing the target, and
intermediate results are used to update the knowledge base when further
reasoning is required. The mathematical representation, reasoning
operations, and success criterion are specified by the chosen proving
method.

\begin{figure}[htb]
    \includegraphics[width=0.7\linewidth]
    {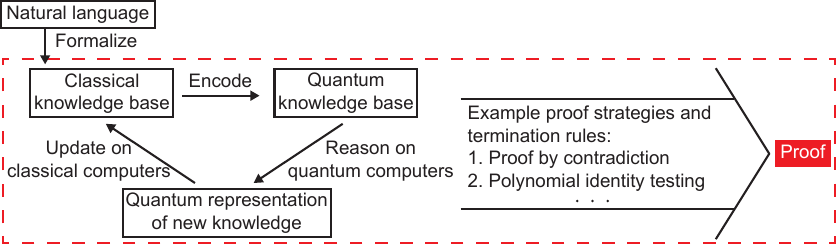}
    \caption{\label{fig-forward-chaining}An iterative QATP workflow with
    classical knowledge-base updates. A natural-language problem is
    formalized into a classical knowledge base, which is encoded as a
    quantum knowledge base. Quantum reasoning produces quantum
    representations of new knowledge. In this scheme, intermediate
    results are post-processed classically to update the knowledge base,
    which is then re-encoded for subsequent reasoning. The selected
    proof strategy and its termination rule determine when the target
    has been established and a proof is obtained. Proof by contradiction
    and polynomial identity testing are shown as examples.}
\end{figure}

Fig.~\ref{fig-forward-chaining} shows the corresponding
classical--quantum workflow. A natural-language problem is first
formalized: the hypotheses define a classical knowledge base, and the
target is expressed in the same formal language. The expressions
required for reasoning are encoded in quantum states or circuits, on
which quantum operations generate the next symbolic result. Examples
include quantum resolution for propositional or first-order formulas
and quantum pseudo-division for polynomials representing geometric
relations. Classical post-processing then updates the representation
used in subsequent steps. The selected proof strategy and its
termination rule determine when this iteration yields a proof.
Depending on the procedure, the computation may also terminate by
refuting the target, when no further progress can be made, or upon
reaching a resource limit.

In Wu's method, the symbolic expressions in this workflow are
polynomials encoding the hypotheses and the target
conclusion~\cite{Wu1978Decision, Chou198809introduction}. The hypotheses
supply polynomial constraints, and the target is represented by a
conclusion polynomial to be reduced. The hypothesis system is first
transformed into triangular form, after which successive pseudo-division
reduces the conclusion polynomial. The pseudo-remainder produced at each
step supplies the polynomial for the next reduction. In the
implementation considered here, classical reconstruction and re-encoding
connect successive reductions, following the update loop in
Fig.~\ref{fig-forward-chaining}. With the variable ordering and
reduction rules fixed, these reductions follow a prescribed algebraic
procedure. The goal test in Fig.~\ref{fig-forward-QATP} then checks
whether the final pseudo-remainder vanishes identically, which
establishes the conclusion under the relevant non-degeneracy conditions.

This final zero test can be formulated as polynomial identity testing:
one searches a suitable hitting set for an evaluation point at which the
pseudo-remainder is nonzero~\cite{Sun2026Quantum}. This connects the
algebraic procedure to the search-based validation tasks studied within
QATP. A related task arises in the quantum resolution procedure
discussed above, where valid resolvents must be identified among
candidate inference results~\cite{Sun2026Quantum}. For these
search-based validation tasks, Grover-type
algorithms~\cite{Grover1996fast, Yoder201411Fixed} can provide a
quadratic reduction in query complexity. This comparison concerns the
respective search subroutines rather than the overall cost of theorem
proving. Superpolynomial advantages in generic settings are constrained
by the relevant no-go results and complexity-theoretic
assumptions~\cite{Kabanets2003Derandomizing,Aaronson2014Need}. For
restricted problem classes with additional structure, such as particular
families of logical formulas or geometric polynomials with specific
symmetries, quantum algorithms may nevertheless admit superpolynomial
speedups beyond the generic quadratic improvement.

Geometry provides a useful setting for demonstrating these algebraic
operations. Statements such as the perpendicularity of the diagonals of
a square are readily understood, and coordinate representations give a
direct formalization in terms of indexed variables. These features make
small geometric examples suitable for noisy intermediate-scale quantum
(NISQ) devices. In this section, we first review classical Wu's
method~\cite{Wu1978Decision, Chou198809introduction}, then present the
square example, and finally describe its quantum implementation and the
simplifications adopted in this work.

\subsection{Classical Wu's method}

Wu's method~\cite{Wu1978Decision, Chou198809introduction} is a
deterministic algebraic framework for establishing consequences of
polynomial equation systems. In geometry theorem proving, each point is
assigned a pair of coordinate variables, and geometric relations, such
as collinearity, parallelism, perpendicularity, and equality of angles,
are translated into multivariate polynomial equations with integer
coefficients. This converts the geometric statement into an algebraic
problem. Let the hypotheses be represented by the finite polynomial set
$\mathcal{H} = \{ H_1, H_2, \dots, H_m \}$ and the conclusion by a
polynomial $G$. The task is to determine whether $G = 0$ follows from
the constraints in $\mathcal{H}$ under generic geometric configurations.

Elimination proceeds with a fixed variable ordering, typically chosen
according to the order in which points are introduced in the geometric
construction. The ordering reflects the constructive structure of the
problem and determines the leading variable and leading coefficient of
each polynomial. Successive pseudo-division steps reduce the polynomial
system according to this ordering and yield a triangular set
$\mathcal{T} = \{ f_1, f_2, \dots, f_k \}$ whose leading variables are
strictly increasing. Together with the prescribed reduction rules, the
variable ordering specifies how the triangularization proceeds.

Consider two polynomials $P$ and $Q$ over a coefficient ring, with the
variable ordering fixed. Let $y$ be the leading variable of $Q$, let
$D_Q = \deg_y(Q)$, and denote the leading coefficient of $Q$ with
respect to $y$ by $L(Q)$. For $\deg_y(P) = D_P \ge D_Q$, pseudo-division
removes the highest power of $y$ in $P$ without dividing by $L(Q)$.
Starting from $P^{(0)} = P$ and $A^{(0)} = 0$, define
$d_i = \deg_y(P^{(i-1)}) - D_Q$ and
$c_i = L\left(P^{(i-1)}\right)$ at each iteration
$i = 1,2,\dots,t$, where $t = D_P-D_Q+1$ and $L(\cdot)$ denotes the
leading coefficient with respect to $y$. The reduction updates the
current polynomial according to
\begin{align}
    P^{(i)} = L(Q)P^{(i-1)} - c_i y^{d_i} Q,
\end{align}
and accumulates the quotient through
\begin{align}
    A^{(i)} = L(Q)A^{(i-1)} + c_iy^{d_i}.
\end{align}
Each reduction cancels the current highest-degree term of $P^{(i-1)}$
in $y$. The result either vanishes or has a strictly lower degree in
$y$. After $t$ iterations, set $R = P^{(t)}$ and $A = A^{(t)}$.
The final pseudo-remainder satisfies $R=0$ or
$\deg_y(R)<\deg_y(Q)$, and the pseudo-division identity reads
\begin{align}\label{eq-remainder}
    L(Q)^t P = A Q + R.
\end{align}
Multiplication by $L(Q)$ at each reduction step avoids rational
coefficients and keeps all intermediate polynomials in the original
coefficient ring.

To make the cancellation explicit, separate the elimination variable $y$
from the remaining variables and write
\begin{align}
    P = \sum\limits_{{\bf{i}},d}^{{\bf{I}},{D_P}}
    p ({\bf{i}},d)\prod\limits_{k = 1}^K
    {x_k^{{i_k}}} {y^d}, \nonumber \\
    Q = \sum\limits_{{\bf{j}},d}^{{\bf{J}},{D_Q}}
    q ({\bf{j}},d)\prod\limits_{k = 1}^K
    {x_k^{{j_k}}} {y^d}.
\end{align}
Here $p(\cdot)$ and $q(\cdot)$ are the coefficients of $P$ and $Q$,
respectively. The multi-indices ${\bf i}=(i_1,\ldots,i_K)$ and
${\bf j}=(j_1,\ldots,j_K)$ specify the degrees of the variables $x_k$,
whose maximal degrees in $P$ and $Q$ are denoted by $I_k$ and $J_k$.
The first reduction step gives
\begin{align}
    {P^{(1)}} = L(Q)P - L(P){y^{{D_P} - {D_Q}}}Q
    = \sum\limits_{{\bf{i}},{\bf{j}},d}^{{\bf{I}},{\bf{J}},{D_P}}
    {p({\bf{i}},d)q({\bf{j}},{D_Q})}
    \prod\limits_{k = 1}^K {x_k^{{i_k} + {j_k}}} {y^d}
    - \sum\limits_{{\bf{i}},{\bf{j}},d}^{{\bf{I}},{\bf{J}},{D_Q}}
    {p({\bf{i}},{D_P})q({\bf{j}},d)}
    \prod\limits_{k = 1}^K {x_k^{{i_k} + {j_k}}}
    {y^{d + {D_P} - {D_Q}}}.
\end{align}
The two contributions to the coefficient of $y^{D_P}$ in $P^{(1)}$
cancel:
\begin{align}
    \sum\limits_{{\bf{i}},{\bf{j}}}^{{\bf{I}},{\bf{J}}}
    {p({\bf{i}},{D_P})q({\bf{j}},{D_Q})}
    \prod\limits_{k = 1}^K {x_k^{{i_k} + {j_k}}}
    - \sum\limits_{{\bf{i}},{\bf{j}}}^{{\bf{I}},{\bf{J}}}
    {p({\bf{i}},{D_P})q({\bf{j}},{D_Q})}
    \prod\limits_{k = 1}^K {x_k^{{i_k} + {j_k}}} = 0,
\end{align}
which cancels the highest-degree term of $P^{(0)}$ in $y$.
Thus, $P^{(1)}$ either vanishes or has a lower degree in $y$
than $P^{(0)}$, ensuring progress of the reduction.

Wu's method applies pseudo-division repeatedly according to the chosen
elimination ordering. The resulting pseudo-remainders reduce the rank of
the system and are retained for use in subsequent eliminations. When no
further elimination is possible, the resulting system forms the
triangular set $\mathcal{T}$. The conclusion polynomial $G$ is then
pseudo-divided successively by the polynomials in $\mathcal{T}$
according to the same ordering. If the final pseudo-remainder vanishes
identically, $G$ belongs to the ideal generated by $\mathcal{H}$ up to
factors arising from leading coefficients, establishing the geometric
conclusion.

The powers of leading coefficients introduced by pseudo-division
generally impose non-degeneracy conditions requiring those coefficients
to be nonzero. These conditions exclude degenerate configurations and
must be handled explicitly in classical Wu's method. Here they are
assumed to hold implicitly and are not treated further, since our focus
is the algebraic structure needed for the quantum implementation.

\subsection{The square example}

\begin{figure}[htb]
    \includegraphics[width=0.2\linewidth]
    {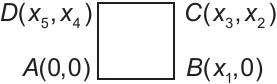}
    \caption{\label{fig-square}Coordinate representation of the square
    $ABCD$ used in Wu's method. Point $A$ is placed at the origin and
    side $AB$ is aligned with the $x$-axis. The remaining vertices are
    parameterized by the algebraic variables $x_1,\dots,x_5$. This
    coordinate choice is made without loss of generality.}
\end{figure}

We now illustrate Wu's method by proving that the diagonals of a square
are perpendicular. The example shows how successive pseudo-division
reduces a target polynomial and how a vanishing final pseudo-remainder
certifies the conclusion. We assume a non-degenerate square $ABCD$ and
choose Cartesian coordinates with $A$ at the origin and $AB$ along the
$x$-axis. The vertices are parameterized as $A = (0,0)$, $B = (x_1,0)$,
$C = (x_3,x_2)$, and $D = (x_5,x_4)$, where $x_1,\dots,x_5$ are
algebraic variables, as illustrated in Fig.~\ref{fig-square}.

The defining properties of the square give four polynomial constraints.
First, ${AB}\perp{AD}$ gives $x_1x_5=0$. Excluding $A=B$ requires
$x_1\neq 0$, so
\begin{align}
H_1: x_5 = 0 .
\end{align}
Second, ${AB}\parallel{CD}$ gives $x_1(x_4-x_2)=0$. Using $x_1\neq 0$
again yields
\begin{align}
H_2: x_4 - x_2 = 0 .
\end{align}
Third, ${AD}\parallel{BC}$ gives $x_5x_2 - x_4(x_3-x_1)=0$.
Substituting $H_1$ reduces this condition to $x_4(x_3-x_1)=0$, and
excluding the degenerate case $x_4=0$ gives
\begin{align}
H_3: x_3 - x_1 = 0 .
\end{align}
Finally, $\lvert AB\rvert=\lvert BC\rvert$ implies
$x_2^2 + (x_3-x_1)^2 - x_1^2 = 0$, which reduces under $H_3$ to
\begin{align}
H_4: x_2^2 - x_1^2 = 0 .
\end{align}
The hypothesis set is therefore
\begin{align}
\mathcal{H} = \{H_1,H_2,H_3,H_4\}.
\end{align}
The desired perpendicularity of $AC$ and $BD$ is encoded by
\begin{align}
G: x_3(x_5-x_1) + x_2x_4 = 0 .
\end{align}

To prove $G=0$ from $\mathcal{H}=\{H_1,H_2,H_3,H_4\}$, we eliminate the
variables in the order $(x_5,x_4,x_3,x_2)$. Starting from $R_0 := G$,
the successive pseudo-remainders are
\begin{align}
R_1 := \mathrm{Prem}(R_0, H_1, x_5), \nonumber \\
R_2 := \mathrm{Prem}(R_1, H_2, x_4), \nonumber \\
R_3 := \mathrm{Prem}(R_2, H_3, x_3), \nonumber \\
R_4 := \mathrm{Prem}(R_3, H_4, x_2).
\end{align}
In the first step, we eliminate $x_5$ in $G$ using $H_1:x_5=0$ by
subtracting $x_3 H_1$:
\begin{align}
R_1= G - x_3 H_1
= (x_3(x_5-x_1)+x_2x_4) - x_3 x_5
= x_2x_4 - x_3x_1 .
\end{align}
Next, subtracting $x_2 H_2$ eliminates $x_4$ using $H_2:x_4-x_2=0$:
\begin{align}
R_2= R_1 - x_2 H_2
= (x_2x_4 - x_3x_1) - x_2(x_4-x_2)
= x_2^2 - x_3x_1 .
\end{align}
The third step eliminates $x_3$ using $H_3:x_3-x_1=0$ by subtracting
$(-x_1)H_3$:
\begin{align}
R_3= R_2 - (-x_1) H_3
= (x_2^2 - x_3x_1) + x_1(x_3-x_1)
= x_2^2 - x_1^2 .
\end{align}
The resulting polynomial $R_3$ coincides with $H_4:x_2^2-x_1^2=0$, so
the final pseudo-remainder vanishes:
\begin{align}
R_4= R_3 - H_4
= (x_2^2 - x_1^2) - (x_2^2 - x_1^2)
= 0 .
\end{align}
Since the final pseudo-remainder is identically zero, Wu's method
certifies that the conclusion polynomial $G$ follows from
$\mathcal{H}=\{H_1,H_2,H_3,H_4\}$ under the non-degeneracy assumptions
used to obtain the hypotheses. The diagonals of the square are
therefore perpendicular.

\subsection{Quantum implementation of Wu's method}

Our quantum implementation represents each polynomial by a circuit
encoding its point-value pairs. This avoids the need to merge like terms
during polynomial multiplication, a central difficulty of
coefficient-based quantum representations~\cite{Sun2026Quantum}.
Polynomial values on a fixed evaluation set are computed classically and
embedded into quantum circuits. The quantum computation acts on these
pointwise values, whereas interpolation and coefficient reconstruction
are performed classically.

For a polynomial $F({\bf x})$ in variables ${\bf x}=(x_1,x_2,\ldots)$,
we choose a finite evaluation set ${{\bf x}_m}$ and calculate
$F({\bf x}_m)$ at each point. A sufficient number of evaluation points
uniquely determines $F$, allowing polynomial operations to be performed
on pointwise data rather than monomial coefficients. The point-value
pairs ${{\bf x}_m, F({\bf x}_m)}$ are encoded by a circuit implementing
\begin{align}
    {U_F}\left| {{\bf{x}_m}} \right\rangle
    \left| {\bf{0}} \right\rangle
    = \left| {{\bf{x}_m}} \right\rangle
    \left| {F({\bf{x}_m})} \right\rangle,
\end{align}
where $\left| {\bf x}_m \right\rangle$ and
$\left| F({\bf x}_m) \right\rangle$ denote the binary encodings of the
evaluation point and its polynomial value. In this representation,
polynomial operations become pointwise reversible integer arithmetic
and require neither term collection nor coefficient comparison.
Circuits for polynomials defined on the same evaluation set can be
composed, making the representation suitable for iterative
pseudo-division.

To use this representation for pseudo-division, we must additionally
express the polynomial as a univariate polynomial in the variable to be
eliminated, say $y$. The coefficients in this representation are
themselves polynomials in the remaining variables. Although this
transformation can in principle be performed by quantum
interpolation~\cite{Sun2026Quantum}, in the present implementation these
coefficient polynomials are obtained classically. They are then encoded
in a quantum circuit whose output is controlled by the exponent $d$ of
$y$. For a polynomial $P$, the resulting coefficient-evaluation circuit
implements
\begin{align}
    {U_{P,y}}\left| d \right\rangle
    \left| {{\bf{x}_m}} \right\rangle
    \left| {\bf{0}} \right\rangle
    = \left| d \right\rangle
    \left| {{\bf{x}_m}} \right\rangle
    \left| {\sum\limits_{\bf{i}}^{\bf{I}}
    p ({\bf{i}},d)\prod\limits_{k = 1}^K
    {x_k^{{i_k}}} } \right\rangle.
\end{align}
Here ${\sum\limits_{\bf{i}}^{\bf{I}}
p ({\bf{i}},d)\prod\limits_{k = 1}^K {x_k^{{i_k}}} }$
is the coefficient of $y^d$ in $P$, regarded as a polynomial in the
remaining variables $\bf{x}$ and evaluated at the point encoded in
$\left| {\bf x}_m \right\rangle$. For the corresponding representation
of $Q$, we abbreviate the output state
$\left| {\sum\limits_{\bf{j}}^{\bf{J}}
q ({\bf{j}},d)\prod\limits_{k = 1}^K
{x_k^{{j_k}}} } \right\rangle $
as $\left| q(d) \right\rangle$ when no ambiguity arises. These
point-value circuits, defined with respect to the chosen elimination
variable, are combined with standard arithmetic
circuits~\cite{RuizPerez201704Quantum} to construct the pseudo-remainder
circuit illustrated in Fig.~2 of the main text.

In the square example, each variable elimination requires only one
pseudo-division step. As shown in the preceding subsection, the first
three steps remove the dependence on $x_5$, $x_4$, and $x_3$,
respectively, while the final step yields the zero polynomial.
Table~\ref{table:mapping} lists the remaining variables $\boldsymbol{z}$, the
elimination variable $y$, and the dividend and divisor $P$ and $Q$ at
each step. These assignments directly connect the algebraic proof to
its quantum circuits.

\begin{table*}[!htbp]
\centering
\begin{tabular}{c|ScScScSc}
\hline
\diagbox{\ \textbf{Variable or Polynomial}\ }{\ \textbf{Step}\ }
&\ \textbf{1}\ \
&\ \textbf{2}\ \
&\ \textbf{3}\ \
&\ \textbf{4}\ \ \\
\hline
$\ \boldsymbol{z}\ $
& $\ (x_{1}, x_{2}, x_{3}, x_{4})\ $
& $\ (x_{1}, x_{2}, x_{3})\ $
& $\ (x_{1}, x_{2})\ $
& $\ (x_{1})\ $ \\
\hline
$\ y\ $
& $\ x_{5}\ $
& $\ x_{4}\ $
& $\ x_{3}\ $
& $\ x_{2}\ $ \\
\hline
$\ P\ $
& $\ G\ $
& $\ x_2x_4 - x_3x_1\ $
& $\ x_2^2 - x_3x_1\ $
& $\ x_2^2 - x_1^2\ $ \\
\hline
$\ Q\ $
& $\ H_{1}\ $
& $\ H_{2}\ $
& $\ H_{3}\ $
& $\ H_{4}\ $ \\
\hline
\end{tabular}

\caption{Variables and polynomials used in the proof.}
\label{table:mapping}
\end{table*}

We implement this four-step elimination on the superconducting quantum
processor using point-value representations throughout. The evaluation
set for each variable is chosen according to its maximal degree at the
corresponding step. In steps 2--4, the highest degree of any individual
variable is two. Where three evaluation values are required, we use
$0$, $1$, and $-1$, as listed in Table~\ref{table:values}. These small
evaluation sets provide sufficient data to determine the relevant
polynomials while limiting the encoding overhead. Together with the
polynomial assignments in Table~\ref{table:mapping}, they specify the
point-value encoding at each elimination step.

\begin{table*}[!htbp]
\centering
\begin{tabular}{c|ScScScSc}
\hline
\diagbox{\ \textbf{Variable}\ }{\ \textbf{Step}\ }
&\ \textbf{1}\ \
&\ \textbf{2}\ \
&\ \textbf{3}\ \
&\ \textbf{4}\ \ \\
\hline
$\ x_{1}\ $
& $\ \{0, 1\}\ $
& $\ \{0, 1\}\ $
& $\ \{0, 1, -1\}\ $
& $\ \{0, 1, -1\}\ $ \\
\hline
$\ x_{2}\ $
& $\ \{0, 1\}\ $
& $\ \{0, 1, -1\}\ $
& $\ \{0, 1, -1\}\ $
& $\ -\ $ \\
\hline
$\ x_{3}\ $
& $\ \{0, 1\}\ $
& $\ \{0, 1\}\ $
& $\ -\ $
& $\ -\ $ \\
\hline
$\ x_{4}\ $
& $\ \{0, 1\}\ $
& $\ -\ $
& $\ -\ $
& $\ -\ $ \\
\hline
$\ x_{5}\ $
& $\ - \ $
& $\ -\ $
& $\ -\ $
& $\ -\ $ \\
\hline
\end{tabular}
\caption{The values of each variable in different reasoning steps.}
\label{table:values}
\end{table*}

In this algebraic realization, the variable ordering and reduction
rules specify the successive reasoning operations. The following
section develops a more general proof-search framework in which
the selection of inference rules and supporting relations is
itself part of the computation, while the corresponding symbolic
transformations are implemented by quantum reasoning circuits.

\section{Quantum proof search}

The preceding section considered a concrete algebraic realization of
QATP, in which the sequence of reasoning operations is specified by the
underlying algebraic procedure. We now turn to a more general setting:
proof search, where the choice of which inference to perform is itself
part of the computational task. Even when individual reasoning
operations can be implemented efficiently, constructing a complete
proof may still require exploring many possible intermediate statements
and reasoning paths~\cite{Sun2026Quantum}. For example, quantum
resolution may involve many rounds of reasoning and repeated searches
for valid resolvents, while algebraic proving methods may generate
increasingly complicated intermediate expressions. Moreover, a proof
obtained through a long sequence of low-level symbolic operations may
be difficult to interpret directly.

\begin{figure}[htb]
    \includegraphics[width=0.6\linewidth]
    {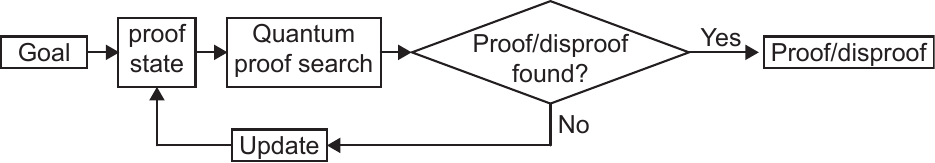}
    \caption{\label{fig-backward-QATP}Schematic quantum proof-search
    loop. Starting from the goal, a proof state is constructed and
    processed by quantum proof search. If a proof or disproof is found,
    the procedure terminates. Otherwise, the proof state is updated and
    the search continues until a proof or disproof is obtained or a
    prescribed stopping condition is met.}
\end{figure}

We now turn to quantum proof search. In the full-angle method considered
here, the proof is constructed through backward chaining: starting from
the target statement, the prover searches for intermediate formulas and
supporting relations that reduce the current proof state toward the
hypotheses. At each step, one must determine which inference rule or 
transformation to apply and,
when necessary, which supporting relations from the knowledge base
should be used. The resulting state then becomes the input to subsequent
reasoning rounds. Fig.~\ref{fig-backward-QATP} illustrates this general
search loop.

The selection of reasoning rounds is a central problem in automated
theorem proving. Classical proof-search systems commonly use
handcrafted heuristics or machine-learning-based strategies to guide
this selection~\cite{Alemi2016DeepMath}. More recently, large language
models have also been explored for generating proofs incrementally or
producing complete proofs
directly~\cite{Collins2024Evaluating,Ren2025DeepSeek,
Hubert2026Olympiad,Zhang2026Proposing}. Such guidance can reduce
unproductive exploration and produce shorter or more interpretable
proofs, although the resulting search procedure need not preserve the
completeness of an exhaustive proving method~\cite{Kaliszyk2018Reinforcement}.

\begin{figure}[htb]
    \includegraphics[width=0.5\linewidth]
    {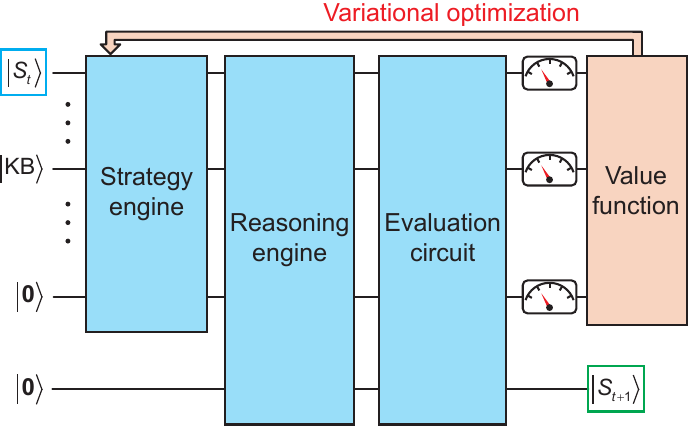}
    \caption{\label{fig-quantum-proof-search-sketch}General quantum
    architecture for proof search. At reasoning round $t$, the current
    proof state $|S_t\rangle$ together with the quantum knowledge base
    $|\mathrm{KB}\rangle$ is supplied to the \textit{strategy engine},
    which generates an action state specifying the reasoning operation
    and supporting relations to be used. The \textit{reasoning engine}
    implements the corresponding formal symbolic transformation and
    produces the next reasoning state $|S_{t+1}\rangle$. The resulting
    state and intermediate information are processed by an
    \textit{evaluation circuit}, which provides signals summarized by
    the \textit{value function}. In the variational implementation
    considered here, these signals are used to update the parameters of
    the strategy circuit. The ellipses indicate possible auxiliary
    qubits used by the strategy and evaluation circuits.}
\end{figure}

Motivated by this perspective, we consider the general quantum
proof-search architecture illustrated in
Fig.~\ref{fig-quantum-proof-search-sketch}. At reasoning round $t$, the
current proof state $|S_t\rangle$ and the quantum knowledge base
$|\mathrm{KB}\rangle$ provide the information used for action
selection. A strategy engine produces an action state that specifies
the inference operation and the supporting relations to be used. The
reasoning engine then implements the corresponding formal symbolic
transformation and produces the next proof state $|S_{t+1}\rangle$.
An evaluation circuit assesses the resulting reasoning state and
provides signals summarized by a value function. These signals can in
turn be used to optimize the strategy engine.

This architecture separates two conceptually different aspects of proof
search. The reasoning engine implements formally specified symbolic
transformations, whereas the strategy engine determines which of these
transformations should be attempted for the current proof state. The
evaluation circuit provides information for improving this selection.
The resulting proof can therefore be checked according to the formal
inference rules independently of how the reasoning actions were chosen.

A quantum implementation also allows symbolic information to remain in
superposition during reasoning. In a fully coherent realization, the
selected actions, supporting relations, and intermediate formulas need
not be read out after every inference step. Instead, their quantum
representations can be retained and processed by subsequent circuits,
allowing different formulas and reasoning paths to coexist within the
same quantum state. This coherent processing provides a basis for
potential quantum advantages in proof search. In the present
experimental implementation, however, the proof proceeds
round-by-round: after each round, the output state is measured and the
selected result is used to prepare the input state for the next round,
as described in the main text.

As a concrete realization of this architecture, we use the full-angle
method~\cite{Chou199612Automated} for automated geometry theorem
proving. Full-angles provide a fixed symbolic representation of
geometric relations, together with a finite set of inference rules and
explicit control strategies for proof construction. These properties
make the method suitable for demonstrating the main components of
quantum proof search on current quantum hardware.

We first introduce the full-angle method and the geometry problem used
in the experiment. We then describe the quantum representation of
formulas, the reasoning engine, and the strategy and evaluation
components. Finally, we present the specific quantum circuits used in
the experimental implementation.

\subsection{Preliminaries for the full-angle method}

The full-angle method~\cite{Chou199612Automated} is a symbolic approach
to automated geometry theorem proving in which geometric relations are
represented and manipulated through \emph{full-angles}. Full-angles are
treated as algebraic objects equipped with equality, addition, and sign
operations, and geometric proofs are constructed by applying a finite
set of inference rules to these objects.

The method combines two complementary procedures. First, a geometry
information base (GIB) is constructed from the hypotheses by deriving
elementary geometric relations that can be used later in the proof.
The target statement is then expressed in terms of full-angles and
reduced by applying inference rules together with the relations stored
in the GIB. A canonical representation and a ranking-based control
strategy restrict the applicable transformations and guide the
reduction toward simpler full-angle expressions. In the terminology of
the classical full-angle method, these two procedures correspond to a
forward construction of the GIB followed by a backward chaining
proof process.

We adopt the following conventions. Points are denoted by uppercase letters, and lines are denoted either by lowercase letters or by pairs of distinct points. The full-angle $\angle[u, v]$ is defined as the angle between the undirected lines $u$ and $v$. Two full-angles $\angle[l, m]$ and $\angle[u, v]$ are equal if there exists a rotation $K$ such that $K(l) \parallel u$ and $K(m) \parallel v$. The addition of two full-angles is defined as $\angle[u, v] + \angle[l, m] = \angle[u, K(m)]$, where $K$ is any rotation satisfying $K(l) \parallel v$.

The full-angles defined as ordered pairs of lines satisfy the following rules~\cite{Chou199612Automated}:
\begin{itemize}
    \item[] \textbf{R1} For all parallel lines $AB \parallel PQ$, $\angle[0] = \angle[AB, PQ]$ is a constant.
    \item[] \textbf{R2} For all perpendicular lines $AB \perp PQ$, $\angle[1] = \angle[AB, PQ]$ is a constant.
    \item[] \textbf{R3} There is an operation ``addition'' between two full-angles which is commutative and associative.
    \item[] \textbf{R4} $\angle[1] + \angle[1] = \angle[0]$.
    \item[] \textbf{R5} $\angle[u, v] + \angle[0] = \angle[u, v]$.
    \item[] \textbf{R6} If $X$ is on line $PQ$, then $\angle[AB, PX] = \angle[AB, PQ]$.
    \item[] \textbf{R7} If $PX$ is parallel to $UV$, then $\angle[AB, PX] = \angle[AB, UV]$.
    \item[] \textbf{R8} If $PX$ is perpendicular to $UV$, then $\angle[AB, PX] = \angle[1] + \angle[AB, UV]$.
    \item[] \textbf{R9} If $XA = XB$ then $\angle[AX, AB] = \angle[AB, XB]$.
    \item[] \textbf{R10} (The Inscribed Angle Theorem) If $A, B, C$, and $D$ are cyclic then $\angle[AD, CD] = \angle[AB, CB]$.
    \item[] \textbf{R11} If $O$ is the circumcenter of triangle $ABC$ and $M$ is the midpoint of $AB$ then $\angle[AO, OM] = \angle[AC, BC]$.
    \item[] \textbf{R12} If $MA = MB$ and $A, B, P, M$ are cyclic then $\angle[PA, PM] = \angle[PM, PB]$.
    \item[] \textbf{R13} $\angle[AB, CD] = -\angle[CD, AB]$.
    \item[] \textbf{R14} For any line $UV$, $\angle[AB, CD] = \angle[AB, UV] + \angle[UV, CD]$.
    \item[] \textbf{R15} If $\text{coll}(A, X, U)$, $\text{coll}(P, X, R)$, and $\text{cyclic}(X, P, Q, U)$, then by R14, R6, and R10 $\angle[AX, BC] = \angle[AX, XP] + \angle[XP, BC] = \angle[QU, QP] + \angle[PR, BC]$.
    \item[] \textbf{R16} If $\text{coll}(X, C, D)$, $\text{coll}(X, U, A)$, $\text{coll}(X, V, B)$, $\text{cyclic}(X, U, C, E)$, and $\text{cyclic}(X, V, D, F)$, then by R14, R6, and R10 $\angle[AX, BX] = \angle[AX, XC] + \angle[XD, XB] = \angle[EU, EC] + \angle[FD, FV]$.
    \item[] \textbf{R17} If $XA = XB$ and $\text{coll}(X, A, W)$ then by R14 and R9 $\angle[AX, BX] = \angle[AX, AB] + \angle[AB, XB] = 2\angle[AX, AB] = 2\angle[AW, AB]$.
    \item[] \textbf{R18} If $XA = XB$, $\text{coll}(X, A, W)$, and $\text{coll}(A, B, U)$ then by R14, R6 and R9 $\angle[BX, CD] = \angle[BX, AB] + \angle[AB, CD] = \angle[AU, AW] + \angle[AU, CD]$.
    \item[] \textbf{R19} If $\text{circumcenter}(O, A, B, C)$ and $\text{midpoint}(M, A, B)$ then by R14, R6 and R11 $\angle[AO, AB] = \angle[AO, OM] + \angle[OM, AB] = \angle[AC, CB] + \angle[1]$.
    \item[] \textbf{R20} If $\text{cyclic}(X, W, U, V)$, $\text{coll}(X, U, A)$, and $\text{perp}(X, V, P, Q)$ then $\angle[AX, BC] = \angle[AX, VX] + \angle[VX, BC] = \angle[WU, WV] + \angle[PQ, BC] + \angle[1]$.
    \item[] \textbf{R21} If $\text{in-center}(I, A, B, C)$ then we have $\angle[AI, AB] + \angle[BI, BC] + \angle[CI, CA] = \angle[1]$.
\end{itemize}

To organize the application of these rules, the full-angle method
introduces a ranking system. We assume a total order on the points,
denoted by ``$<$''. A full-angle $\angle[AB,CD]$ is in canonical form
if $A>B$, $C\ge D$, and either $A>C$ or
($A=C$ and $B>D$). In the following, all full-angles are assumed to be
written in canonical form.

The ranking between two full-angles is defined lexicographically.
We write
$\angle[AB,CD] < \angle[PQ,RS]$
if one of the following conditions holds:
(1) $A<P$;
(2) $A=P$ and $B<Q$;
(3) $A=P$, $B=Q$, and $C<R$; or
(4) $A=P$, $B=Q$, $C=R$, and $D<S$.

Based on this ranking, two control strategies restrict the application
of inference rules:
\begin{mdframed}
\textbf{Control Strategy 1:}
A rule $\mathcal{R}$ can be applied to $\angle[AB,CD]$ only if
$\text{rank}(\angle[AB,CD])
>
\text{rank}(\mathcal{R}(\angle[AB,CD]))$.
\end{mdframed}

\begin{mdframed}
\textbf{Control Strategy 2:}
Rule R14 (line splitting) can be applied to $\angle[AB,CD]$ only if the
resulting two full-angles can both be further reduced to lower rank.
\end{mdframed}

The first strategy requires each accepted transformation to lower the
rank of the full-angle being reduced. The second places an additional
restriction on R14, which introduces an intermediate line and replaces
one full-angle by two. It permits this splitting operation only when
both resulting terms admit further reductions. Together, these
conditions suppress unproductive transformations and limit unnecessary
growth of the proof-search space.

Before the goal-directed proof begins, the prover constructs the GIB
from the hypotheses. The GIB stores elementary geometric facts that can
subsequently be used to instantiate and simplify the full-angle
inference rules. For the problem considered below, the GIB is generated
as follows:
\begin{enumerate}
    \item \textbf{Initialization:} Add the hypotheses to the GIB:
    \item \textbf{Apply forward chaining rules to derive new information}:
    \begin{itemize}
        \item \textbf{F1}: If $l_1 \parallel l_2$ and $l_1 \parallel l_3$ then $l_2 \parallel l_3$. 
        \item \textbf{F2}: If $l_1 \perp l_2$ then $l_1 \perp l_3$ iff $l_2 \parallel l_3$.
        \item \textbf{F3}: $AB$ is the mediator of $XY$ if and only if $AX = AY$ and $BX = BY$.
        \item \textbf{F4}: If $\text{midpoint}(O, C, A)$ then $AB \perp BC$ iff $\text{circumcenter}(O, A, B, C)$.
        \item \textbf{F5}: If $PA \perp PB$ then $QA \perp QB$ iff $\text{cyclic}(A, B, P, Q)$. 
        \item \textbf{F6}: If $M$ and $N$ are the midpoints of AB and AC then $MN \parallel BC$. 
        \item \textbf{F7}: If $\angle[OU, OP] = \angle[OP,OV] \neq \angle[0]$, $OU \perp PU$, and $OV \perp PV$ then $OU = OV$ and $PU = PV$.
        \item \textbf{F8}: If $AB \parallel AC$ then $\text{coll}(A, B, C).$
    \end{itemize}
    \item \textbf{Iterate:} Continue applying the forward chaining rules to the newly derived information until no new facts can be added to the GIB.
    \item \textbf{Final GIB:} The GIB now contains the original hypotheses and the newly deduced information. This will be used by the prover during the backward chaining process to simplify and eliminate full-angles.
\end{enumerate}

For problems that require additional geometric information, the
following auxiliary rules can also be used:
\begin{itemize}
    \item \textbf{K1:} $l_1 \parallel l_2$ iff there is another line $l_3$ such that $\angle[l_1, l_3] = \angle[l_2, l_3]$.
    \item \textbf{K2:} If $l_1 \perp l_2$ and $l_3 \perp l_4$ then $\angle[l_1, l_3] = \angle[l_2, l_4]$.
    \item \textbf{K3:} $A, B, C,$ and $D$ are cyclic iff $\angle[AB, AC] = \angle[DB, DC]$. (The non-degenerate condition is $\neg \text{coll}(A, B, C, D)$.)
    \item \textbf{K4:} $OA = OB$ iff $\angle[OA, AB] = \angle[AB, OB]$. (The non-degenerate condition is $\neg \text{coll}(O, A, B)$.)
    \item \textbf{K5:} If $MA = MB$ and $A, B, P, M$ are cyclic then $\angle[PA, PM] = \angle[PM, PB]$.
\end{itemize}

\subsection{IMO example}

\begin{figure}[htb]
		\includegraphics[width=0.2\linewidth]{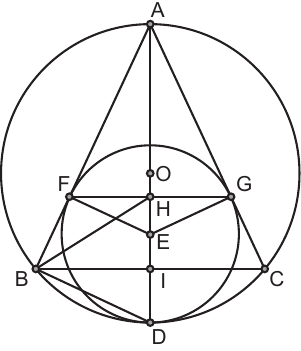}
		\caption{\label{fig-imo1978}Schematic illustration of the IMO 1978 geometry problem.}
    \end{figure}

We use the following geometry problem as a concrete instance for the
quantum proof-search demonstration. In triangle $ABC$, $AB = AC$. A
circle is tangent to the circumcircle of triangle $ABC$ and is tangent
to $AB$ and $AC$ at points $F$ and $G$, respectively. Show that the
midpoint of segment $FG$ is the incenter of triangle $ABC$. The
geometric configuration is shown in Fig.~\ref{fig-imo1978}.

\textbf{Point Order:} $A, B, C, D, E, F, G, H$

\textbf{Hypotheses:}
\begin{itemize}
    \item $\text{cyclic}(A, B, C, D)$ (Points $A, B, C, D$ lie on a circle)
    \item $\text{cong}(A, B, A, C)$ ($AB = AC$)
    \item $\text{circumcenter}(O, A, B, C)$ (Point $O$ is circumcenter of $ABC$)
    \item $\text{foot}(F, E, A, B)$ ($EF \perp AB$)
    \item $\text{foot}(G, E, A, C)$ ($EG \perp AC$)
    \item $\text{cong}(E, F, E, G)$ ($EF = EG$)
    \item $\text{cong}(E, F, E, D)$ ($EF = ED$)
    \item $\text{coll}(E, O, D)$ (Points $E, O, D$ collinear)
    \item $\text{cong}(F, H, H, G)$ ($FH = HG$)
    \item $\text{coll}(F, G, H)$ (Points $F, G, H$ collinear)
\end{itemize}

\textbf{Conclusion:}
\begin{itemize}
    \item $\angle HBA + \angle HBC$ = 0 (Line $BH$ bisects angle $ABC$)
\end{itemize}

Before performing the full-angle proof search, we construct the GIB by
deriving additional geometric relations from the hypotheses. For the
present problem, the following relations are obtained:
\begin{itemize}
    \item $\text{cyclic}(A, E, F, G)$. Proof: $EF \perp FA$, $EG \perp GA$ (F5).
    \item $AF=AG$. Proof: $EF = EG \to \angle FAE = \angle EAG \to \angle GEA = \angle AEF \to AF = AG$ (reverse of K5).
    \item $\text{coll}(A, H, E)$, $AE \perp FG$. Proof: $AF=AG$, $EF=EG$ (F3).
    \item $BD=CD$. Proof: $\angle BAD = \angle DAC$ (reverse of K5).
    \item $BI=IC$, $AD \perp BC$. Proof: $AB=AC$, $BD=CD$ (F3).
    \item $AO \perp BC$. Proof: $AB=AC$, $BO=CO$ (F3).
    \item $\text{coll}(A, O, E)$. Proof: $AO \perp BC$, $AD \perp BC$ (F2, F8).
    \item $FG \parallel BC$. Proof: $\angle [AD, BC] = \angle [1] = \angle[AD, FG]$. (F2).
    \item $EF \parallel BD$. Proof: $\angle BDA = \angle BCA = \angle FGA = \angle FEA$. (K1, K5).
    \item $AB \perp BD$. Proof: $EF \parallel BD$, $EF \perp AB$. (F2).
    \item $\text{cyclic}(F, H, D, B)$. Proof: $FH \perp HD$, $FB \perp BD$. (F5).
\end{itemize}

Together with the original hypotheses, these derived relations form the
geometric information used in the subsequent full-angle reasoning. To
simplify the GIB representation, parallel, collinear, and perpendicular
relations are used to replace equivalent line representations by those
with lower rank. The relevant geometric relations in this problem are
$CB \parallel GF$, $FE \parallel DB$,
$\text{coll}(A, O, H, E, I, D)$,
$\text{coll}(F, G, H)$, $\text{coll}(B, C, I)$,
$EF \perp AB$, $EG \perp AC$, $AD \perp BC$, and
$AB \perp BD$.
This simplification allows equivalent full-angles to be represented in
lower-ranked forms. For example, $\angle[HG, FE]$ is written as
$\angle[HG, DB]$ because $FE \parallel DB$ and
$\text{coll}(F, G, H)$. Similarly, $\angle[HG, FE]$ can be expressed
as $\angle[HG, BA] + \angle[1]$ using $EF \perp AB$.

For this example, we use the basic rules R9, R10, and R12 to establish
the angle relations. The corresponding geometric relations are:

\begin{enumerate}
    \item \textbf{R9:} $AB=AC$, $EF=EG$, $EF=ED$, $AF=AG$, $BD=CD$.
    \item \textbf{R10:} $\text{cyclic}(A, B, C, D)$; $\text{cyclic}(A, E, F, G)$; $\text{cyclic}(F, H, D, B)$.
    \item \textbf{R12:} $AB=AC$ and $\text{cyclic}(A, B, C, D)$; $EF=EG$ and $\text{cyclic}(A, E, F, G)$; $AF=AG$ and $\text{cyclic}(A, E, F, G)$; $BD=CD$ and $\text{cyclic}(A, B, C, D)$.
\end{enumerate} 

The corresponding angle relations are:
\begin{enumerate}
    \item \textbf{R9:} $ \angle ABC = \angle BCA$, $ \angle EFG = \angle FGE$, $ \angle EFD = \angle FDE$, $ \angle AFG = \angle FGA$, $ \angle DBC = \angle BCD$.
    \item \textbf{R10:} $\angle ACB = \angle ADB$, $\angle ABC = \angle ADC$, $\angle ABD = \angle ACD$, $\angle BAC = \angle BDC$, $\angle BAD = \angle BCD$, $\angle CAD = \angle CBD$; 
    $\angle AFE = \angle AGE$, $\angle AEF = \angle AGF$, $ \angle AEG = \angle AFG$, $\angle EAF = \angle EGF$, $\angle EAG = \angle EFG$, $\angle FAG = \angle FEG$;
    $\angle FDH = \angle FBH$, $\angle FHD = \angle FBD$, $ \angle FHB = \angle FDB$, $\angle HFD = \angle HBD$, $\angle HFB = \angle HDB$, $\angle DFB = \angle DHB$.
    \item \textbf{R12:} $\angle ADB = \angle CDA$, $\angle EAF = \angle GAE$,$ \angle AEF = \angle GEA$, $\angle DAB = \angle CAD$.
\end{enumerate} 

These relations provide the geometric information from which the
proof-search actions used later are constructed. In the following
subsections, we describe how the corresponding formulas and proof
states are represented on a quantum computer and how the reasoning
operations act on them.

\subsection{State representation of formulas}

The quantum proof-search framework operates on quantum representations
of symbolic formulas. Two types of states play distinct roles. The
\emph{proof state} $|S_t\rangle$ represents the formula, or
superposition of formulas, being processed at reasoning round $t$,
whereas the \emph{knowledge-base state} stores the hypotheses and other
relations that may be used to support subsequent inferences. In our
previous work~\cite{Sun2026Quantum}, we showed that a knowledge base in
conjunctive normal form (CNF) for propositional logic can be encoded as
a quantum superposition in which each basis state represents a clause.
Other symbolic objects can be represented in a similar manner. In the
full-angle method considered here, the relevant formulas are
full-angles~\cite{Chou199612Automated}.

At the beginning of the goal-directed search, the target formula is
encoded as the initial proof state. For a classically specified target,
this state can be prepared directly as a computational-basis product
state. Each subsequent reasoning operation acts on $|S_t\rangle$ and
produces the state $|S_{t+1}\rangle$. When successive reasoning circuits
are composed coherently, the circuit depth required to prepare
$|S_t\rangle$ increases linearly with the number of reasoning rounds
$t$, provided that the depth of each reasoning operation is bounded as
discussed below.

The knowledge base provides the hypotheses and supporting relations
used by these reasoning operations. If the knowledge base is specified
classically, it can be loaded onto a quantum computer using an index
register, following the encoding introduced in
Ref.~\cite{Sun2026Quantum}. The situation is more general when new
formulas derived during reasoning are also incorporated into the
knowledge base. In that case, the updated knowledge base may either be
constructed classically and re-encoded, or generated directly as a
quantum state.

For simple instances in which only a small number of formulas are
added, classical updating is sufficient. For larger instances, one may
instead retain the derived formulas coherently. Consider an extreme
case in which every result produced at a reasoning round is new. If a
knowledge base containing $M$ formulas is combined with another copy of
itself, pairwise reasoning produces up to $M^2$ derived formulas. If
this enlarged knowledge base is used in the same way at the next step,
the number of represented formulas can grow to $M^4$. Continuing this
construction for $t$ reasoning rounds gives a state representing up to
$M^{2^t}$ formulas, while the circuit depth of the composed reasoning
operations grows linearly with $t$.

Alternatively, the initial knowledge base
$\left| {\rm{KB}}_0 \right\rangle$ containing $M$ formulas can be
combined at every step with the current knowledge-base state
$\left| {\rm{KB}}_t \right\rangle$ to produce
$\left| {\rm{KB}}_{t+1} \right\rangle$. Under the corresponding
counting assumption, the number of represented formulas then grows as
$M^t$ rather than $M^{2^t}$. The circuit depth remains linear in $t$,
while the number of index qubits required to label the formulas changes
from $2^t\log_2 M$ in the first construction to $t\log_2 M$ in the
second.

These estimates concern the number of formulas represented within a
quantum state rather than the number of formulas extracted as classical
outputs. Explicitly storing or reasoning over $M^{2^t}$ or $M^t$
formulas as separate classical records would require correspondingly
large classical memory. A quantum state, by contrast, can retain these
formula labels in superposition. This does not imply an exponential
quantum advantage for producing the same formulas explicitly in
classical form. As discussed in Ref.~\cite{Sun2026Quantum}, extracting
a large knowledge base by measurement and sampling removes this
representational separation. The distinction is therefore between
maintaining a coherent quantum representation for subsequent reasoning
and materializing the entire set of formulas as a classical data
structure.

The knowledge base used in the full-angle implementation is the
geometry information base (GIB). It contains elementary and easily
derived geometric relations~\cite{Chou199612Automated}. In the present
experiment, the GIB is constructed classically using the geometric rules
introduced above and is then encoded for use by the quantum reasoning
circuits. This classical preprocessing substantially simplifies the
subsequent backward proof
search~\cite{Nevins1975Plane}. More general implementations could
instead generate or update parts of the supporting knowledge coherently,
without reading out all intermediate formulas.

To address formulas coherently within a large knowledge base, each
formula is associated with an index register. For a knowledge base
loaded directly from classical data, the encoding can be written as
\begin{align}
    {U_\text{KB}}\left| i \right\rangle \left| {\bf{0}} \right\rangle  = \left| i \right\rangle \left| {{F_i}} \right\rangle,
\end{align}
where $U_{\text{KB}}$ denotes the knowledge-base encoding unitary,
$i$ is the formula index, $\left| {\bf{0}} \right\rangle$ is the
initial state of the data register, and $F_i$ is the $i$-th formula in
the knowledge base.
If a formula is generated coherently from two indexed premises, its
index can be formed from the premise indices themselves. A reasoning
operation can be written as
\begin{align}
    {U_{\rm{R}}}\left| i \right\rangle \left| j \right\rangle \left| {{F_i}} \right\rangle \left| {{F_j}} \right\rangle \left| {\bf{0}} \right\rangle  = \left| i \right\rangle \left| j \right\rangle \left| {{F_i}} \right\rangle \left| {{F_j}} \right\rangle \left| {R({F_i},{F_j})} \right\rangle,
\end{align}
where $U_{\rm{R}}$ denotes the forward chaining reasoning circuit, and the pair $(i,j)$ serves as the index of the derived formula $R(F_i, F_j)$.

When formulas are generated under action selection, their indices can
similarly incorporate both the index of the supporting formula and the
selected action. The exact index structure depends on the architecture
of the strategy engine. For example, the strategy engine may process
$|S_t\rangle\otimes|\mathrm{KB}_t\rangle$ or, when two supporting
relations are required,
$|S_t\rangle\otimes|\mathrm{KB}_t\rangle
\otimes|\mathrm{KB}_t\rangle$.
Different choices change the information available for action selection
and lead to different requirements for circuit design and optimization.
A systematic analysis of these alternatives is beyond the scope of the
present work.

The index register specifies which formula is being addressed, while a
separate data register encodes the formula itself. The number of index
qubits grows logarithmically with the number of represented formulas.
The cost of the data register instead depends on the complexity of an
individual formula. In the representation used here, a formula with
$N$ variables requires $N$ qudits, with each qudit having a local value
set of cardinality $|V|$. A qubit implementation therefore requires
$N\log_2|V|$ qubits for the formula data.

For example, the CNF encoding introduced in
Ref.~\cite{Sun2026Quantum} uses two qubits for each propositional
variable. The four basis states $|00\rangle$, $|01\rangle$,
$|10\rangle$, and $|11\rangle$ represent absence, positive, negative,
and resolved status, respectively. In the full-angle representation
used here, a formula $\pm\angle[DB,CA]$ is encoded as
$\left|\pm\right\rangle\left|D\right\rangle
\left|B\right\rangle\left|C\right\rangle\left|A\right\rangle$.
The first register represents the sign, while the remaining registers
encode the point indices. For instance,
$-\angle[DB,CA]$ is represented by the basis state
$\left|13120\right\rangle$, where the first digit denotes the negative
sign and the remaining digits encode the ranks of $D$, $B$, $C$, and
$A$, respectively.

With these representations, the proof state, supporting knowledge, and
formula indices can all be processed directly by quantum circuits. The
next subsection describes the reasoning engine that implements the
formal transformations between such encoded formulas.

\subsection{Reasoning engine}

The reasoning engine shown in
Fig.~\ref{fig-quantum-proof-search-sketch} serves as the logical
inference module in the QATP framework. It implements the formal
inference rules of the underlying logic system and transforms the
formulas at one reasoning round into those at the next step according to
the selected inference rule.

The reasoning circuit takes three types of input: (i) the quantum state
representing the formulas at the current reasoning round, (ii) an action
state specifying which inference rule is applied and which hypotheses
from the knowledge base are used, and (iii) the quantum-encoded
knowledge base containing the available hypotheses. The structure of
the reasoning engine directly reflects the inference rules of the logic
system, and its concrete implementation depends on the chosen encoding
of symbolic formulas.

A key feature of the quantum proof-search framework is that the
reasoning engine can operate coherently on formulas represented in
superposition. Quantum parallelism allows the same inference procedure
to act simultaneously on the different formulas encoded in the input
state. For the examples considered below, the corresponding reasoning
circuits have constant depth or a depth that grows only mildly with the
cardinality of the local value set. Consequently, even when the proof
state represents a superposition of many formulas, the inference
operation can be carried out without processing these formulas one by
one.

We first recall the quantum resolution circuit introduced in
Ref.~\cite{Sun2026Quantum}. Each propositional variable can take one of
four local values: absence, positive, negative, and resolved. The
resolution circuit acts on all propositional variables in parallel and
resolves literals with opposite polarity in the two premise clauses.
As a result, the circuit depth of the resolution operation itself does
not increase with the number of propositional variables.

As a second example, we consider rule \textbf{R9} of the full-angle
method. This example illustrates how a full-angle inference rule can be
implemented as a quantum reasoning circuit. We assume that the strategy
engine has selected rule \textbf{R9} and specified the hypotheses to be
used. Under this assumption, the reasoning engine applies the
corresponding inference to the input formula state.

Consider a knowledge base containing three indexed geometric relations:
00:$AC = AB$, 01:$FE = FD$, and 10:$GF = GA$. In the full-angle
method, this knowledge base, or geometry information base (GIB), is
encoded as a unitary operation that loads the corresponding full-angle
representations into the data register conditioned on the action state:
\begin{align}
    {U_{{\rm{KB}}}}\left| {00} \right\rangle \left| {\bf{0}} \right\rangle  = \left| {00} \right\rangle \left| { - \angle [{CB},{BA}]} \right\rangle \left| {\angle [{CB},{CA}]} \right\rangle, \nonumber \\
    {U_{{\rm{KB}}}}\left| {01} \right\rangle \left| {\bf{0}} \right\rangle  = \left| {01} \right\rangle \left| { - \angle [{FD},{ED}]} \right\rangle \left| {\angle [{FE},{ED}]} \right\rangle, \nonumber \\
    {U_{{\rm{KB}}}}\left| {10} \right\rangle \left| {\bf{0}} \right\rangle  = \left| {10} \right\rangle \left| { - \angle [{GA},{FA}]} \right\rangle \left| {\angle [{GF},{FA}]} \right\rangle.
\end{align}
The first line corresponds to the equation $\angle [CB, CA] = -\angle [CB, BA]$, which encodes the geometric relation $AC = AB$ under rule \textbf{R9}. Here we consider only the forward application of the rule, since the control strategy requires that the rank of the formula be reduced during inference; analogous considerations apply to the other relations. The unitary $U_{\rm{KB}}$ can be obtained by direct encoding, brute-force forward chaining, or forward chaining guided by the strategy engine. The translation between geometric relations (e.g., $AC = AB$) and full-angle relations (e.g., $\angle[CB,CA] = -\angle[CB,BA]$) under a given rule is straightforward and is not the focus here. The reasoning task is to reason on an unknown input state $\left|\pm\angle[ZX,YW]\right\rangle$, where points $Z,X,Y,W$ may coincide.

\begin{figure}[htb]
		\includegraphics[width=0.85\linewidth]{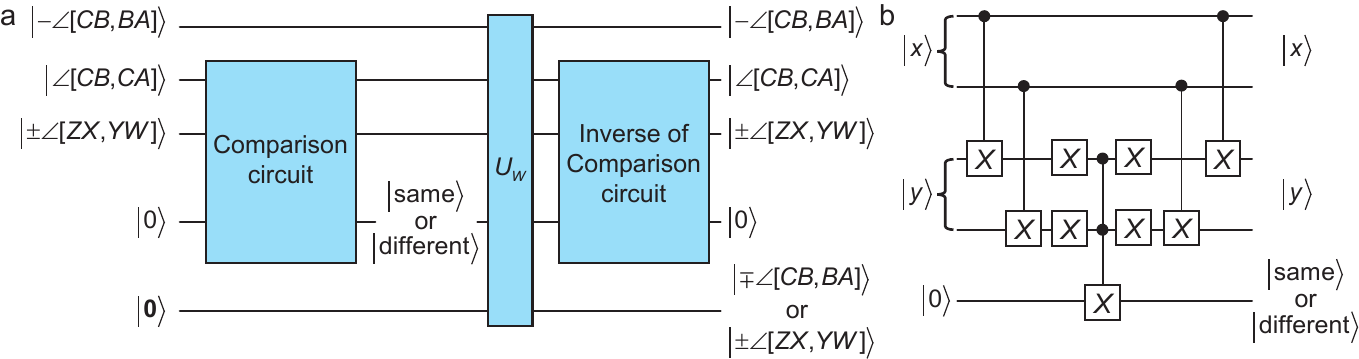}
		\caption{\label{fig-angle-reason} Quantum circuit implementing rule \textbf{R9}. \textbf{a}, Reasoning circuit for the input state $ \left| \pm \angle [ZX, YW] \right\rangle$ based on the rule equation $\angle [CB, CA] = -\angle [CB, BA]$. The circuit first compares the input state $\left| \pm  \angle [ZX, YW] \right\rangle$ with the premise of the rule equation. If the comparison yields the state $\left| {{\rm{same}}} \right\rangle $, namely $[ZX, YW]=[CB, CA]$, the circuit $U_W$ outputs the conclusion $ \left| \mp \angle [CB, BA] \right\rangle$; otherwise, it leaves the input state unchanged. \textbf{b}, Circuit for comparing two $n$-qubit registers $\left|x\right\rangle$ and $\left|y\right\rangle$. In this illustration, each register contains two qubits. The circuit compares the two registers qubit by qubit. The CX gates between corresponding qubits of $\left|x\right\rangle$ and $\left|y\right\rangle$ map each qubit in the $\left|y\right\rangle$ register to $\left|0\right\rangle$ when the corresponding qubits are the same and to $\left|1\right\rangle$ when they are different. The following $X$ gates convert the equality condition into controls for an $n$-controlled NOT gate, which writes the comparison result to an ancillary qubit. The final gates uncompute the comparison operation and restore the input registers. For two $n$-qubit registers, the circuit contains $2n$ CX gates arranged in two parallel layers, $2n$ single-qubit $X$ gates, and one $n$-controlled NOT gate. The latter can be implemented either with depth $O(\log_2 n)$ using $O(n)$ auxiliary qubits, or directly using $O(n)$ CX gates~\cite{Barenco199511Elementary}. The ancillary output is $\left|1\right\rangle$ when the two registers are the same and $\left|0\right\rangle$ when they are different.
        }
\end{figure}

We first consider the case in which the action state is $\left|00\right\rangle$, corresponding to the relation $AC = AB$. The circuit shown in Fig.~\ref{fig-angle-reason}\textbf{a} compares the input state $\left|\pm\angle[ZX,YW]\right\rangle$ with the state $\left|\angle[CB,CA]\right\rangle$ using auxiliary qubits. If the comparison indicates equality, the output register is prepared in the state $\left|\mp\angle[CB,BA]\right\rangle$; otherwise, the original input state $\left|\pm\angle[ZX,YW]\right\rangle$ is preserved. After the operation, all auxiliary qubits are reset. The unitary $U_W$ shown in Fig.~\ref{fig-angle-reason} implements a controlled copy operation conditioned on the comparison outcome. When the outcome is ``same'', $U_W$ writes the conclusion state to the output register and flips the sign qubit accordingly; when the outcome is ``different'', it copies the original input state to the output register.

If the action state is instead a superposition of $\left|00\right\rangle$, $\left|01\right\rangle$, and $\left|10\right\rangle$, the same reasoning circuit remains valid. In this case, the output state becomes entangled with the action index, such that inference is performed with $AC=AB$, $FE=FD$, and $GF=GA$ conditioned on the respective action states. Moreover, if the input formula state is itself a superposition, the output is a superposition of the corresponding inference results. In other words, the reasoning circuit implements rule \textbf{R9} simultaneously for all input formulas and all geometric relations in the GIB by exploiting quantum parallelism.

Circuits for other inference rules can be constructed in a similar manner. Consequently, the action state produced by the strategy engine naturally contains two indices: one specifying which inference rule is applied, and the other specifying which hypothesis from the knowledge base is used. The circuit depth of the reasoning engine scales as $O(|R|\log|V|)$, where $|R|$ is the number of inference rules and $|V|$ is the cardinality of the value set of variables. The linear dependence on $|R|$ arises because inference rules are applied sequentially, while the logarithmic dependence on $|V|$ follows from the decomposition of multi-qubit controlled operations, as illustrated in Fig.~\ref{fig-angle-reason}\textbf{b} and in the quantum resolution circuit of Ref.~\cite{Sun2026Quantum}. For the full-angle method considered here, $|R|=21$, corresponding to the 21 classical full-angle inference rules, and $|V|=8$, corresponding to the eight possible points $A,B,\ldots,H$ in the example of Fig.~\ref{fig-imo1978}. For propositional QATP in Ref.~\cite{Sun2026Quantum}, $|R|=1$ (resolution) and $|V|=4$, corresponding to the four possible literal states: absence, positive, negative, and resolved.

More broadly, this construction reflects a general pattern in symbolic reasoning: an inference step typically consists of checking whether the input matches the applicability pattern of a rule, and then performing the prescribed transformation once the match is confirmed. For this reason, the circuit design presented here is not limited to resolution in propositional logic or to the full-angle method, but applies more generally to a broad class of rule-based reasoning systems, thereby illustrating the generality of the QATP framework.

Finally, we emphasize that the number of variables $N$—or, in the full-angle setting, the number of points or full-angles appearing in a formula—does not affect the depth of the reasoning circuit. This is because inference on different variables can be performed in parallel. Tasks such as verifying whether a reasoning result is valid (e.g., checking that exactly one variable is resolved in quantum resolution) are not handled by the reasoning engine itself. Instead, these tasks are delegated to the strategy engine and the value function evaluation, which together guide the selection of inference rules and hypotheses and enforce the validity of the reasoning process.

\subsection{Strategy engine and value function}

Within the proof-search framework, the strategy engine determines the
reasoning actions to be considered, while the evaluation circuit
assesses the resulting reasoning states and provides information for
improving action selection. In the implementation considered here, the
strategy engine is realized as a variational quantum circuit whose
parameters determine the distribution over candidate actions. The
evaluation circuit provides signals that are used to optimize these
parameters. We do not attempt to give a comprehensive analysis of the
expressive power of the strategy engine, the optimization landscape, or
the complexity of circuit compilation. Instead, we present two
heuristic scenarios to illustrate how quantum superposition and
interference may be useful for action selection and strategy
evaluation. These examples are intended as illustrative constructions
rather than formal complexity-theoretic guarantees. More general
training schemes, including reinforcement-learning-based optimization
over multi-step proof trajectories, are left for future work.

The first scenario considers a simple setting with three possible formulas, ${F_1}:\left|00\right\rangle$, ${F_2}:\left|01\right\rangle$, and ${F_3}:\left|10\right\rangle$, together with two available reasoning actions. Action $\left|0\right\rangle$ derives $F_3$ from either $F_1$ or $F_2$, whereas action $\left|1\right\rangle$ maps $F_1$ to $F_2$ and $F_2$ to $F_1$. For the input formula $F_1$, both actions are reasonable, since action $\left|0\right\rangle$ produces the new formula $F_3$ and action $\left|1\right\rangle$ produces the new formula $F_2$. The same holds for the input formula $F_2$. It is therefore natural for the strategy engine to assign equal weight to the two actions in both cases. A Hadamard gate acting on the second qubit realizes exactly this behavior: it prepares the action state $(\left|0\right\rangle+\left|1\right\rangle)/\sqrt{2}$ for $F_1$ and $(\left|0\right\rangle-\left|1\right\rangle)/\sqrt{2}$ for $F_2$.

The same gate also yields appropriate behavior for superposed inputs. In particular, when the input is $(\left|F_1\right\rangle+\left|F_2\right\rangle)/\sqrt{2}$, the amplitude of action $\left|1\right\rangle$ vanishes by destructive interference, so that only action $\left|0\right\rangle$ remains. This is desirable because action $1$ merely exchanges $F_1$ and $F_2$ and therefore produces no new formula for this superposed input, whereas action $0$ generates the new formula $F_3$. More generally, a single Hadamard gate already provides a suitable strategy for the three relevant inputs $F_1$, $F_2$, and $(\left|F_1\right\rangle+\left|F_2\right\rangle)/\sqrt{2}$. This example illustrates that even a very simple quantum strategy engine can adapt coherently to different reasoning states and automatically suppress actions that are unproductive for the current superposed input. By contrast, a classical strategy engine would generally need to query the input formulas separately in order to determine which formulas are present and whether a given action can generate new information.

The second scenario is inspired by the Deutsch-Jozsa algorithm~\cite{Deutsch199212Rapid}. Suppose that we need to evaluate the performance of two strategy engines. The first is completely ineffective: under the actions generated by this strategy, every reasoning output coincides with its input. For an input state $\sum\nolimits_{m=1}^{M}\left|F_m\right\rangle/\sqrt{M}$, the resulting output state is $\sum\nolimits_{m=1}^{M}\left|F_m\right\rangle\left|F_m\right\rangle/\sqrt{M}$. The second strategy engine is more effective: for half of the possible inputs, the reasoning result differs from the input. For the same input superposition, the output state becomes $\sum\nolimits_{m=1}^{M}\left|F_m\right\rangle\left|F_{f(m)}\right\rangle/\sqrt{M}$, where $f(m)=m$ holds for exactly half of the values of $m$. After applying a comparison circuit between the input and output registers and storing the comparison result in an auxiliary qubit, where the comparison circuit is of the type shown in Fig.~\ref{fig-angle-reason}\textbf{b}, distinguishing these two cases reduces to a task analogous to that addressed by the Deutsch-Jozsa algorithm. In this sense, the value-evaluation circuit can exploit quantum parallelism to assess whether a strategy engine is entirely unproductive or generates genuinely new reasoning outcomes for half of the all possible input formulas.

We emphasize that these scenarios are deliberately constructed examples and may not reflect the typical situations encountered during the actual learning process of the quantum proof-search system. We also acknowledge that a classical algorithm could efficiently solve these tasks if granted direct access to the relevant oracle information. The purpose of these examples is therefore not to claim a rigorously established exponential quantum advantage in a standard classical input-output setting, but rather to illustrate the possibility that certain tasks within QATP may benefit from quantum strategy engines and quantum value-evaluation circuits. In particular, when quantum states or quantum circuits are provided as inputs and coherent quantum states are retained throughout the computation, superposition states can represent exponentially many formulas and action paths simultaneously, and quantum interference between different reasoning paths can be exploited~\cite{Stahlke201408Quantum}. This suggests that, in suitably structured instances of quantum automated theorem proving, quantum advantages analogous to those found in algorithms based on the quantum Fourier transform (QFT)~\cite{Coppersmith2002approximate,Shor1994Algorithms}, the Deutsch-Jozsa algorithm~\cite{Deutsch199212Rapid}, and Simon's algorithm~\cite{Simon199710Power} may also emerge.

\subsection{Real circuit}

\subsubsection{Circuit of reasoning engine}

In our experiment, we encode lines rather than points in order to reduce the number of qubits required to represent full-angles on a quantum computer. To further reduce the qubit cost and circuit depth, we exclude parallel and collinear lines from the encoding. Under this convention, the first component of a full-angle involves three possible lines, encoded as HD:00, HB:01, and FD:10, while the second component involves four possible lines, encoded as FD:00, DB:01, BC:10, and BA:11.

At the level of circuit design, there is a tradeoff between generality and implementation overhead. Following the general construction of the reasoning circuit introduced above, one may regard each inference rule, such as \textbf{R9}, \textbf{R10}, or \textbf{R12}, as a distinct strategy, and then build a reasoning circuit that matches the input formula against both the selected inference rule and the geometric relation in the GIB to which the rule is applied. An advantage of this approach is that the circuit depth does not increase with the number of geometric relations stored in the GIB, since the matching can be carried out coherently in parallel. However, for the present experiment on superconducting processors, such a construction introduces substantial overhead and is difficult to implement reliably.

We therefore adopt a more concrete compilation scheme. Instead of treating an entire inference rule as a single strategy, we treat each explicit equality between full-angles obtained from applying an inference rule to a specific geometric relation in the GIB as an independent strategy. Under this scheme, the corresponding circuits are much easier to design and implement. The cost is that the number of available actions increases with the number of explicit equalities included in the compiled reasoning scheme, and the circuit depth grows accordingly. Nevertheless, when both the GIB and the set of inference rules are small, this overhead remains acceptable for near-term devices.

Importantly, this simplification preserves the coherent action of the
reasoning circuit on superposed proof states and action states. For a
superposition of candidate inputs, the compiled circuit applies the
corresponding symbolic transformations coherently to all components to
which the selected actions are applicable. If multiple reasoning rounds
are composed coherently without intermediate measurements, the proof
state can in principle contain a rapidly growing number of reasoning
branches, potentially representing an exponentially large space of
reasoning paths. In the present experiment, however, the proof search is
implemented round by round: the output is measured after each round and
the selected result is used to prepare the input state for the next
round. Thus, the experiment does not directly realize coherent
multi-round growth of the proof-path superposition. We nevertheless
adopt the compiled scheme because it retains the coherent symbolic
reasoning within each quantum circuit while substantially reducing the
hardware overhead. The relevant equalities used in the present proof
are listed below:

\begin{enumerate}
    \item \textbf{R9:} $\angle [HD, FD] = \angle [FD, DB]$
    \item \textbf{R10:} $\angle [HB, BA] = \angle [HD, FD]$, $\angle [FD, DB] = -\angle [HB, BC]$
\end{enumerate} 
In the line-index representation, these relations become
\begin{enumerate}
	\item \textbf{R9:} $\angle [00, 00] = \angle [10, 01]$
	\item \textbf{R10:} $\angle [01, 11] = \angle [00, 00]$, $\angle [10, 01] = -\angle [01, 10]$
\end{enumerate} 
For convenience, we also write the proof sequence explicitly:
$\angle [HB,BA] + \angle [HB,CB] = \angle [HD,FD] + \angle [HB,CB] = \angle [FD,DB] + \angle [HB,CB] =  - \angle [HB,CB] + \angle [HB,CB] = \angle 0$.
In the line-index representation, this proof sequence becomes $\angle [1, 3] = \angle [0, 0] = \angle [2, 1] = -\angle [1, 2]$.

\begin{figure}[htb]
		\includegraphics[width=0.7\linewidth]{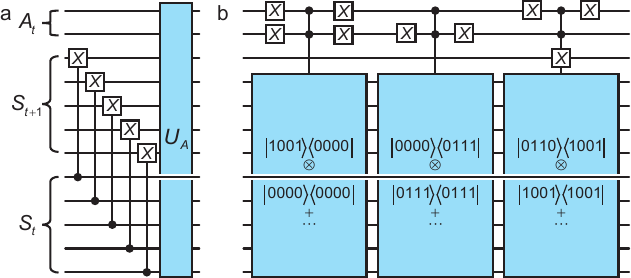}
		\caption{\label{fig-reason-circuit-exp}Quantum reasoning circuit for IMO example. \textbf{a}, Overall structure of the 12-qubit circuit. The first two qubits encode the action $A_t$, the next five qubits represent the proof state $S_{t+1}$ at reasoning round $t+1$, and the final five qubits represent the proof state $S_t$ at reasoning round $t$. The CNOT gates first copy the computational-basis encoding of $S_t$ to the $S_{t+1}$ register, so that the state is preserved when no effective action is applied. Then the action-controlled unitary $U_A$ is performed on $A_t$ and $S_{t+1}$. \textbf{b}, Decomposition of $U_A$ into action-conditioned mapping blocks. Each block implements the angle-state transformation associated with one action, including both positive and negative versions of the relevant full-angle relation. The terms explicitly shown correspond to the transitions used in the present proof, while the omitted terms complete the unitary on the remaining basis states and leave unchanged those states to which no effective action applies.
        }
\end{figure}

The core part of each reasoning round uses 12 qubits as shown in Fig.~\ref{fig-reason-circuit-exp}. The first two qubits encode the action index, the next five qubits represent the proof state $S_{t + 1}$ at reasoning round $t+1$, and the final five qubits represent the proof state $S_t$ at reasoning round $t$. Before applying the reasoning operation, we first copy the proof state $S_t$ to the middle five qubits. This initialization ensures that the state remains unchanged when no effective action is applied. We then implement the unitary $U_A$, shown in Fig.~\ref{fig-reason-circuit-exp}\textbf{b}, which performs an action-controlled update of the state $S_{t+1}$ conditioned on both the action index and the matched component of $S_t$:
\begin{align}\label{eq-unitary-action}
U_A = & \left| {00} \right\rangle \left\langle {00} \right| \otimes I \otimes \left| {1001} \right\rangle \left\langle {0000} \right| \otimes I \otimes \left| {0000} \right\rangle \left\langle {0000} \right| \nonumber \\
+ & \left| {01} \right\rangle \left\langle {01} \right| \otimes I \otimes \left| {0000} \right\rangle \left\langle {0111} \right| \otimes I \otimes \left| {0111} \right\rangle \left\langle {0111} \right| \nonumber \\
+ & \left| {10} \right\rangle \left\langle {10} \right| \otimes X \otimes \left| {0110} \right\rangle \left\langle {1001} \right| \otimes I \otimes \left| {1001} \right\rangle \left\langle {1001} \right| \nonumber \\
+  & \cdots.
\end{align}

More specifically, when the action qubits are fixed to a given basis state, $U_A$ updates the $S_{t+1}$ register according to the selected action, provided that the corresponding component in $S_t$ matches the required input relation. For example, when the action register is $\left|00\right\rangle$, the unitary maps the component $\left|0000\right\rangle$ in $S_{t+1}$ to $\left|1001\right\rangle$, conditioned on the matched component $\left|0000\right\rangle$ in $S_t$, thereby implementing the transition $\angle[HD,FD]\to\angle[FD,DB]$. The terms shown in Eq.~(\ref{eq-unitary-action}) are the transitions used in the present proof, while the omitted terms complete the unitary on the remaining basis states and leave unchanged those states to which no effective action applies.

\subsubsection{Circuit for the value function}

In the general proof-search framework, the evaluation circuit provides
signals that characterize the result of a reasoning action and can be
used to guide the optimization of the strategy engine. In the present
implementation, we use a simple evaluation criterion based on a
coherent comparison between the proof states at two consecutive
reasoning rounds, $t$ and $t+1$. This criterion favors actions that
produce a nontrivial update of the proof state and suppresses an unused
action state in the two-qubit action register. More general evaluation
functions, including those that account for longer reasoning
trajectories, are left for future work.

\begin{figure}[htb]
		\includegraphics[width=0.25\linewidth]{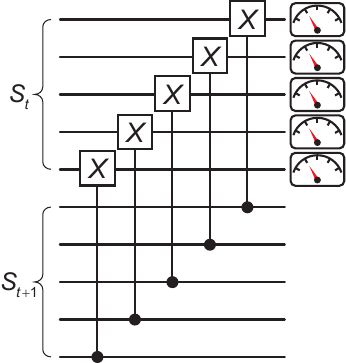}
		\caption{\label{fig-circuit-value}The quantum circuit to measure $P_1$ in Eq. (\ref{eq-value}). This circuit coherently compares the formulas at steps $t$ and $t+1$. The measurement result is ``00000'' when the corresponding components in the two proof states are identical. As an example, consider the input state $3/5\left| {00110} \right\rangle  \otimes \left| {00110} \right\rangle  - 4/5\left| {00010} \right\rangle  \otimes \left| {10100} \right\rangle $, which represents the superposition of two reasoning processes: one in which the formula ``00110''($ + \angle [HB,BC]$) remains unchanged, and another in which the formula ``10100''($ - \angle [HB,FD]$) is derived from ``00010'' ($ + \angle [HD,BC]$). After this circuit, the state evolves to $3/5\left| {00000} \right\rangle  \otimes \left| {00110} \right\rangle  - 4/5\left| {10110} \right\rangle  \otimes \left| {10100} \right\rangle $. The $\left| {00000} \right\rangle $ on the first five qubits indicates that the corresponding proof-state component is unchanged. Sampling the measurement outcomes yields $P_1 = 9/25$. By optimizing the action circuit to reduce this probability, the fraction of reasoning actions that produce no new formulas can be suppressed.
        }
\end{figure}

The loss function used in the experiment is constructed from two measurement-derived contributions,
\begin{align}\label{eq-value}
L = P_1 + P_2 .
\end{align}
Both terms are extracted from the joint measurement statistics according to
\begin{align}
P_1
&=
\Pr!\left(S_t\oplus\widetilde{S}{t+1}=00000\right),\
P_2
&=
\Pr!\left(A_t=11,S_t\oplus\widetilde{S}{t+1}\neq00000\right).
\end{align}
The first contribution collects the total probability of joint measurement events in which the two proof-state registers return the same outcome. As illustrated in Fig.~\ref{fig-circuit-value}, five CNOT gates map the bitwise difference of the two proof-state outcomes onto the comparison register, for which the outcome ``00000'' corresponds to equal results. The second contribution collects events associated with the unused action outcome ``11'' only when the two proof-state measurement outcomes are different, so that events already included in $P_1$ are not counted again. The remaining three action outcomes, 00'', 01'', and ``10'', correspond to the three implemented reasoning actions.

\subsubsection{Circuit of strategy engine}

\begin{figure}[htb]
		\includegraphics[width=0.7\linewidth]{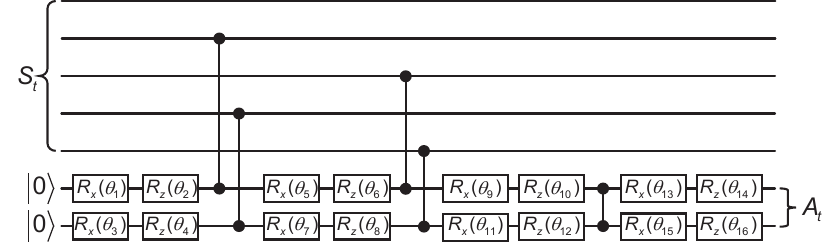}
		\caption{\label{fig-circuit-strategy} Variational circuit used to determine the action register $A_t$ at reasoning round $t$. In the encoding of the proof state $S_t$, the first qubit represents the sign and the remaining four qubits encode the indices of the two lines in the full-angle. Each qubit in the action register is entangled with the line-index qubits in $S_t$, while single-qubit rotation gates are inserted before and after the two-qubit entangling gates to enhance the expressive power of the circuit. The circuit contains 16 variational parameters.
        }
\end{figure}

We use a variational circuit to determine the action index at each reasoning round. In this work, the action at the $t$-th step, denoted by $A_t$, is determined solely by the proof state at the same step, $S_t$. The circuit is designed so that each qubit in the action register interacts with the qubits in $S_t$ that encode the geometric content relevant for action selection. To enhance the expressive power of the ansatz, each two-qubit entangling gate is accompanied by single-qubit rotation gates. This hardware-efficient design allows the circuit to represent a rich family of action-selection policies. In future extensions, more expressive circuits may be employed to decide actions based on multiple copies of formulas in the current proof state as well as information stored in the GIB.

\section{Experiment}
\subsection{Device information}
\begin{figure*}[!htbp]
    \includegraphics[height=0.9\textheight]{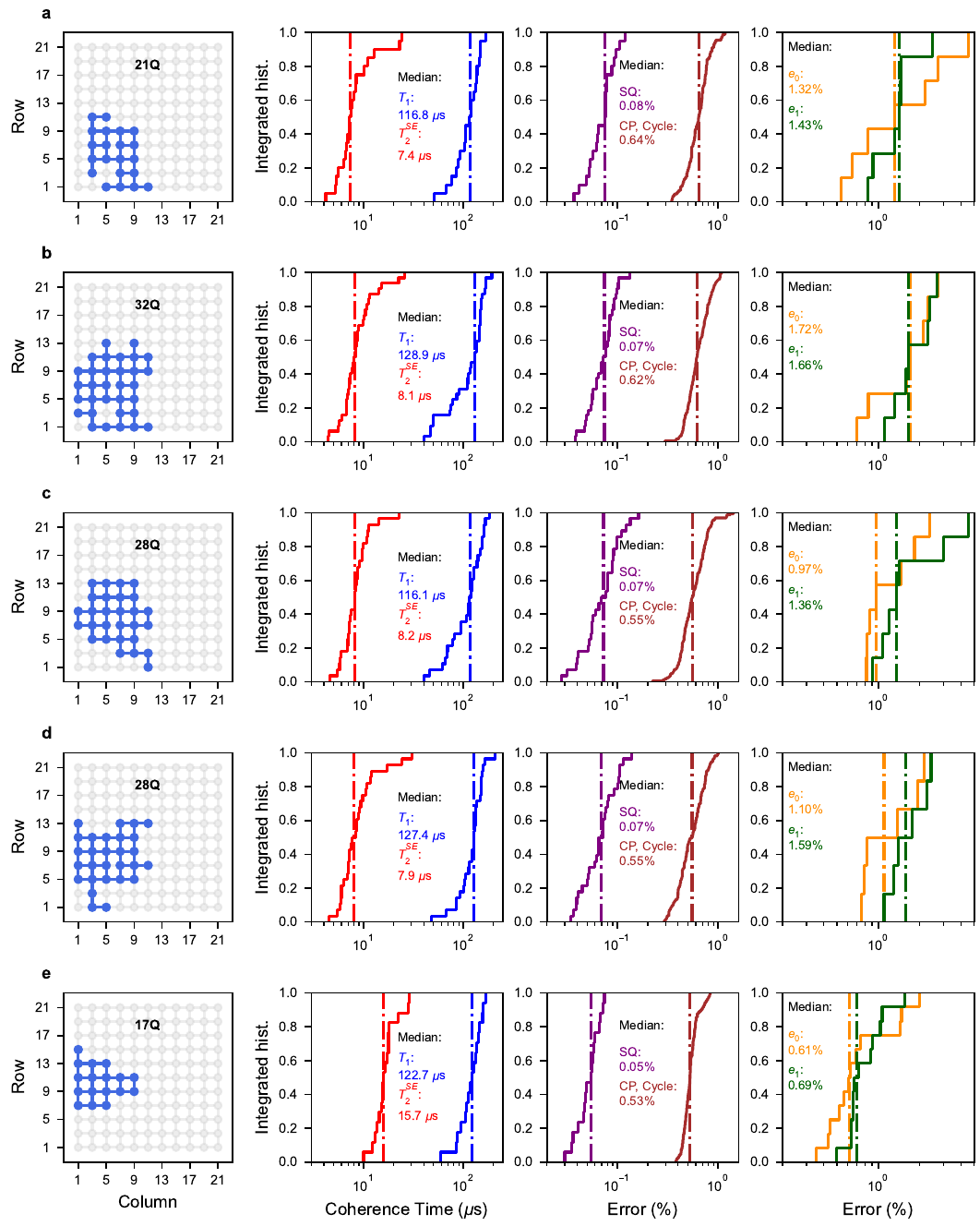}
    \caption{\textbf{Device layout and performance for different quantum circuits.}
        \textbf{a-e}, Device layout and integrated histograms of device parameters. \textbf{Panels in the first column:} Blue dots (bars) denote active qubits (couplers) which are selected for different quantum circuits. The numbers of qubits are 21, 32, 28, 28, 17, respectively. \textbf{Panels in the second column:} Cumulative distributions of qubit energy relaxation time $T_{1}$ (blue line) and spin echo dephasing time $T_{2}^{SE}$ (red line). \textbf{Panels in the third column:} Pauli errors of simultaneous single-qubit gates (purple line) and cycle Pauli errors of two-qubit $CP$ gates (brown line). \textbf{Panels in the fourth column:} Readout errors of state $|0\rangle$ (orange line) and $|1\rangle$ (green line). Vertical dashed lines indicate the median values.\label{fig:Perf.}
    }
\end{figure*}

\begin{figure*}[!htbp]
    \includegraphics[width=\textwidth]{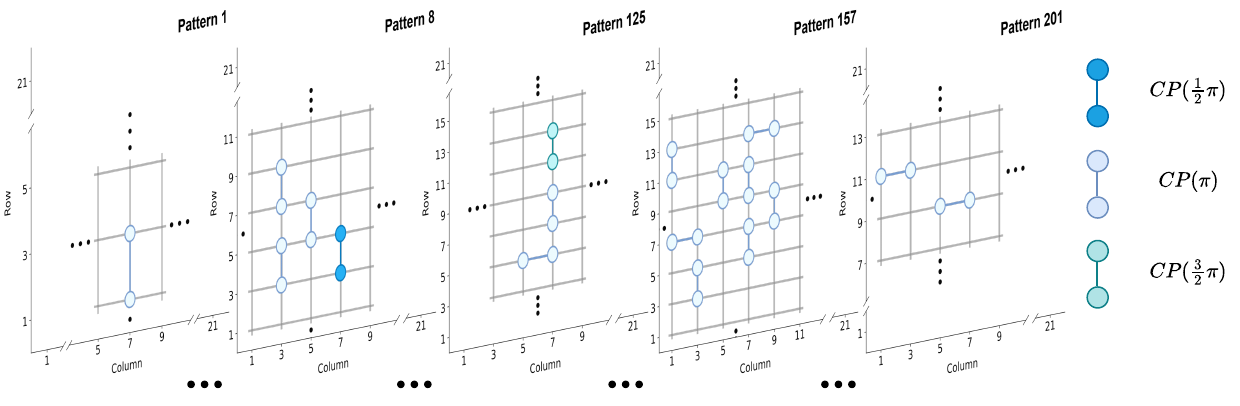}
    \caption{\textbf{Patterns of parallel $CP$ gates.} 
    The gray chip layout is shown as a visual reference, while colored circles and lines indicate the active qubits and couplers, respectively, involved in implementing the $CP$ gates. There are three types of $CP$ gates parameterized by different control phases. We group them into separate patterns according to whether they can be executed without conflicts. The gate count and type may vary across different patterns. In this work, the total number of patterns is 201.}
    \label{fig:Patterns}
\end{figure*}

The superconducting quantum processor utilized in this work employs an 11$\times$11 array of frequency-tunable transmon qubits arranged on a two-dimensional square lattice. In addition, the processor includes 220 couplers mediating nearest-neighbor interaction and 121 resonators for qubit state readout. All quantum reasoning circuits are realized through single-qubit, two-qubit, and measurement gates.
\begin{enumerate}[label=-]

    \item \textbf{Single-qubit gates.} Arbitrary single-qubit gates are implemented through a combination of microwave $XY$ rotations and virtual $Z$ rotations. All microwave pulses employ Gaussian envelopes with a duration of 24 ns and utilize DRAG modulation to mitigate leakage errors during fast gate operations~\cite{chen2016measuring}. We assess single-qubit gate errors via simultaneous cross-entropy benchmarking (XEB). A summary of single-qubit parameters, such as the energy relaxation times $T_1$, the spin-echo dephasing times $T_2^{SE}$ and gate errors, is displayed in the second and third columns of Fig.~\ref{fig:Perf.}. It should be noted that all coherence times are measured simultaneously at the qubit idle frequencies.
    \item \textbf{Two-qubit gates.} The specific two-qubit gate employed in the experiments is the controlled-phase ($CP$) gate. The $CP$ gate with a control phase $\phi$ can be defined as 
    \begin{align}
        CP(\phi) = \begin{pmatrix}
        1 & 0 & 0 & 0\\
        0 & 1 & 0 & 0\\
        0 & 0 & 1 & 0\\
        0 & 0 & 0 & e^{i\phi}\\
        \end{pmatrix}.
    \end{align}
    The realization of the $CP$ gate relies on tuning the states $|11\rangle$ and $|20\rangle$ (or $|02\rangle$) near resonance and activating qubit-qubit coupling to accumulate a $\phi$ (where $\phi \in \{\frac{\pi}{2}, \pi, \frac{3\pi}{2}\}$ in this work) phase shift. This process completes in 44 ns. The detailed calibration methods can be found in ref.~\cite{Jin2025Topological}. As shown in Fig.~\ref{fig:Patterns}, we perform parallel $CP$ gates based on a variety of distinct patterns. We assess gate errors by performing simultaneous XEB, in which we execute circuits comprising multiple cycles; each cycle consists of two random single-qubit $\pi/2$ rotations on both qubits followed by one $CP$ gate. The XEB cycle errors are shown in the third column of Fig.~\ref{fig:Perf.}.
    \item \textbf{Measurement gates.} Our quantum processor utilizes a frequency multiplexed dispersive readout scheme. There are 16 readout lines each of which couples to multiple qubit-resonator pairs. Measurement gates are implemented through microwave pulses with custom-designed envelopes. We optimize the readout pulse power, readout pulse length, and qubit frequency during readout to maximize measurement performance. Furthermore, to enhance readout fidelity, we apply an additional microwave pulse to drive the transition of  $|1\rangle \leftrightarrow |2\rangle$ prior to the readout pulse. Readout errors, displayed in the fourth column of Fig.~\ref{fig:Perf.}, are quantified by preparing random bitstring quantum states and calculating the probability of misclassified states.
    
\end{enumerate}

\subsection{Circuit and layout}
In this section, we introduce details associated with the quantum circuits executed in our experiments. 
To execute the theoretical circuits with our quantum processor, we apply two pre-processing procedures including multi-qubit gate optimization and quantum circuit compilation.

\begin{figure*}[!htbp]
    \centering
    \includegraphics[width=\textwidth]{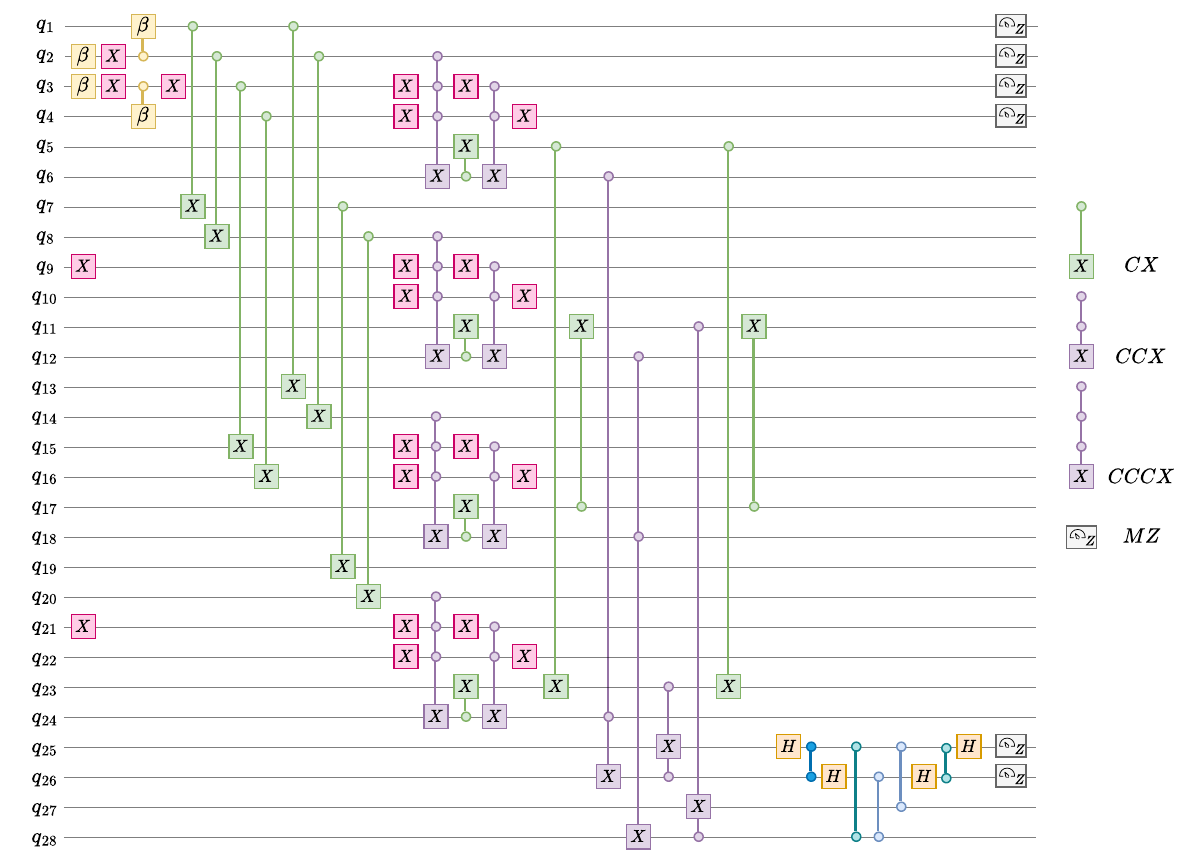}
    \caption{\textbf{The theoretical quantum circuit of step 4 executed in the algebraic theorem proving.} This circuit involves 28 logical qubits, multiple types of quantum gates and 6 necessary $Z$ basis measurement gates for the result extraction. Here, $\beta$ = $\text{arccos}\ \frac{1}{3}$.}
        \label{fig:square_4_theo.}
\end{figure*}
\subsubsection{Multi-qubit gate optimization}
Multi-qubit controlled gates are important building blocks of quantum circuits, enabling complex conditional operations. Unlike single-qubit rotations and several two-qubit gates, however, direct and high-fidelity implementations of multi-qubit controlled gates remain challenging on most quantum processors, particularly for arbitrary groups of qubits. Therefore, high-level multi-qubit operations must be decomposed into sequences of single- and two-qubit gates for execution on quantum processors. Such decompositions must carefully account for the noise and hardware constraints of noisy intermediate-scale quantum (NISQ) devices.

\begin{figure*}[!htbp]
        \centering
        \includegraphics[width=\textwidth]{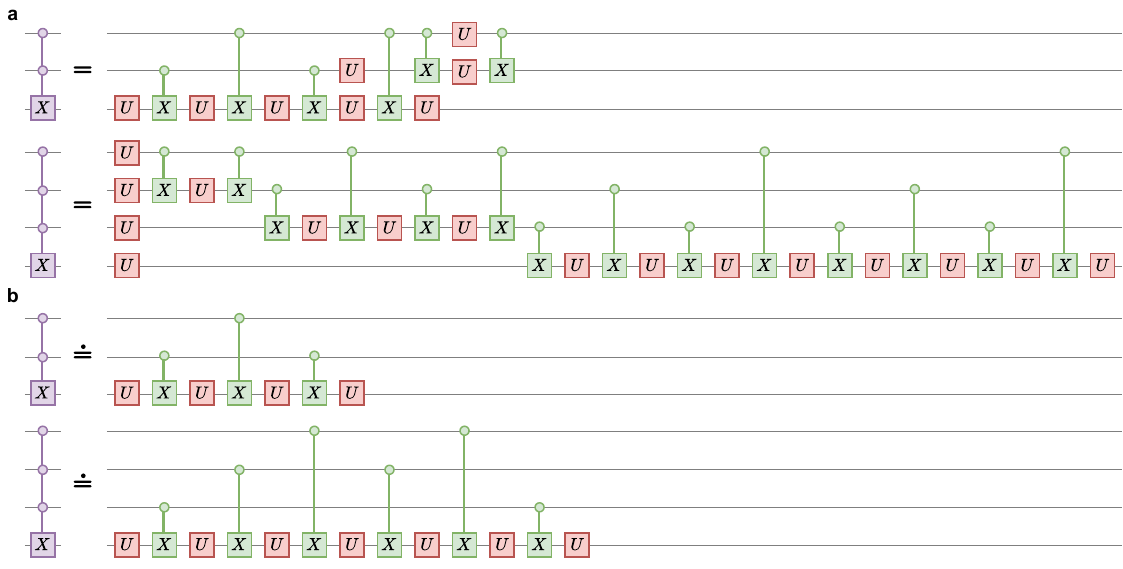}
        \caption{\textbf{The original (a) and optimized (b) implementations of the $CCX$ and $CCCX$ gates.} It should be noted that the decomposition of $CCX$ and $CCCX$ gates into sequences of $U$ and $CX$ gates is not unique. However, the minimal number of required $CX$ gates is the same. For notational simplicity, we ignore the parameters of different $U$ gates. }
        \label{fig:MQgates}
\end{figure*}

As an example, the theoretical quantum circuit shown in Fig.~\ref{fig:square_4_theo.} contains several multi-qubit gates, such as controlled-controlled-X ($CCX$) and controlled-controlled-controlled-X ($CCCX$) gates. Decomposing them into the universal gate set $\{U, CX\}$ generates long gate sequences (see Fig.~\ref{fig:MQgates}\textbf{a}). The arbitrary single-qubit $U$ and two-qubit $CX$ gates can be expressed as 
\begin{align}
    U(\theta, \varphi, \lambda) = \begin{pmatrix}
    \mathrm{cos}\ \frac{\theta}{2} & -e^{i\lambda}\mathrm{sin}\ \frac{\theta}{2} \\
    e^{i\varphi}\mathrm{sin}\ \frac{\theta}{2} & e^{i(\varphi+\lambda)}\mathrm{cos}\ \frac{\theta}{2} \\
    \end{pmatrix}
\end{align}
and 
\begin{align}
    CX = \begin{pmatrix}
    1 & 0 & 0 & 0\\
    0 & 1 & 0 & 0\\
    0 & 0 & 0 & 1\\
    0 & 0 & 1 & 0\\
    \end{pmatrix}.
\end{align}
 Such sequences not only increase circuit overhead, but also exacerbate decoherence effects, which degrade the fidelity of circuit results. To maximally suppress these detrimental impacts while preserving circuit validity, we adopt optimized implementations of the $CCX$ and $CCCX$ gates up to a relative phase~\cite{maslov2016advantages}. As shown in Fig.~\ref{fig:MQgates}\textbf{b}, we achieve an approximate 50$\%$ reduction in both quantum gate count and circuit depth relative to the original implementation. We remark that reduction in the number of gates, especially error-prone two-qubit gates, is particularly valuable as they are one of the most significant error sources in current NISQ devices.

\subsubsection{Quantum circuit compilation} 
Quantum circuit compilation is the critical translation layer that bridges theoretical quantum circuits and physical hardware. This process involves mapping and routing qubits for limited connectivity, transpiling circuit with native gate set and scheduling gates to minimize idle times and crosstalk.

\begin{figure*}[!htbp]
        \centering
        \includegraphics[width=\textwidth]{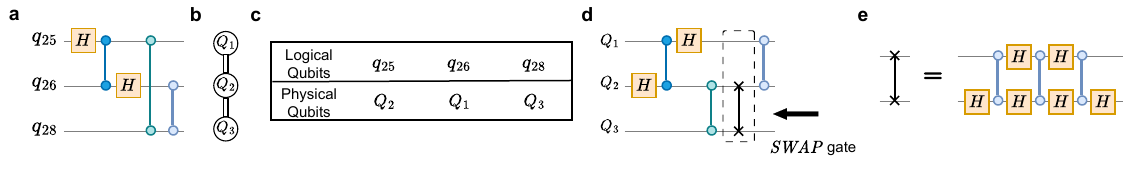}
        \caption{\textbf{A simple example for qubit mapping and routing.} 
        \textbf{a}, A piece of the theoretical quantum circuit shown in Fig.~\ref{fig:square_4_theo.}.
        \textbf{b}, A coupling layout for a chain consisting of three physical qubits.
        \textbf{c}, The table of qubit mapping relations.
        \textbf{d}, The transformed circuit corresponds to the circuit in \textbf{a}. We insert a {\it SWAP} gate to reposition $Q_2$ and $Q_3$ for the third $CP$ gate.
        \textbf{e}, A decomposition of {\it SWAP} gate.
        }
        \label{fig:QMR}
\end{figure*}

With several two-qubit gates requiring long range interactions, the theoretical quantum circuits after optimization can still not be executed with our superconducting quantum processor featuring the nearest-neighbor coupling architecture. To address this constraint, we adopt SWAP-based bidirectional heuristic algorithm (SABRE) to dynamically map logical qubits in the theoretical circuits to physical qubits on the quantum processor based on actual coupling layout~\cite{li2019tackling}.
We prioritize assignments that minimize long-range two-qubit gates and use {\it SWAP} gates for unavoidable non-adjacent interactions. For instance, a theoretical circuit piece (Fig.~\ref{fig:QMR}\textbf{a}) needed to be executed utilizing three physical qubits $Q_1, Q_2, Q_3$ on our superconducting quantum processor. The coupling layout for these physical qubits is depicted in Fig.~\ref{fig:QMR}\textbf{b}. Suppose that we map logical and physical qubits according to the relations listed in the Table (Fig.~\ref{fig:QMR}\textbf{c}). The first two $CP$ gates can be executed directly. But for the third, we must insert a {\it SWAP} gate to route the states of $Q_2$ and $Q_3$, which enables the necessary interactions. The transformed circuit is shown in Fig.~\ref{fig:QMR}\textbf{d}.

Following the initial transformation, the circuits are transpiled into our quantum processor's native gate set $\{U, CP\}$. During this, we also apply a gate fusion optimization: whenever a few single-qubit gates act consecutively on the same qubit, we merge them into one equivalent gate through unitary matrix multiplication. This fusion step further reduces the total gate count and shortens the circuit depth. Next, we schedule the single-qubit and $CP$ gates into separate, temporally isolated layers to minimize the effect of crosstalk. Furthermore, we rearrange them as close in time as possible to minimize qubit idle times between dependent gate operations, which suppresses the impact of decoherence.

\subsubsection{Experimental circuits}
\begin{figure*}[!htbp]
        \centering
        \includegraphics[width=\textwidth]{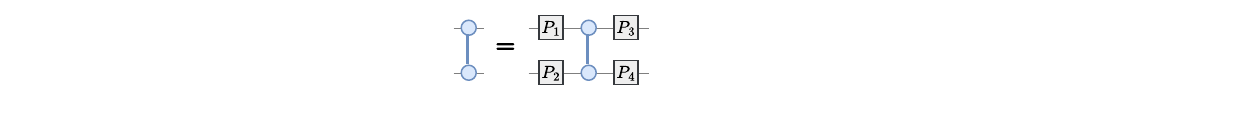}
        \caption{Equivalent circuit illustrating the Pauli Twirling. $P_{i}\ (i=1, 2, 3, 4)$ is a Pauli gate chosen to leave the $CP(\pi)$ gate invariant. For example, $P_1$ = $P_3$ = $Y$, $P_2$ = $I$, $P_4$ = $Z$.
        }
        \label{fig:PT}
\end{figure*}

Beyond the pre-processing procedures described above, we additionally apply Pauli Twirling to four experimental circuits executed in the algebraic theorem proving. The Pauli twirling technique can suppress gate-dependent errors (e.g. coherent over-rotation or dephasing) by sandwiching the $CP(\pi)$ gates with 16 random Pauli gate combinations (see Fig.~\ref{fig:PT}), leaving the overall function invariant.
    
\begin{figure*}[!htbp]
        \centering
        \includegraphics[width=\textwidth]{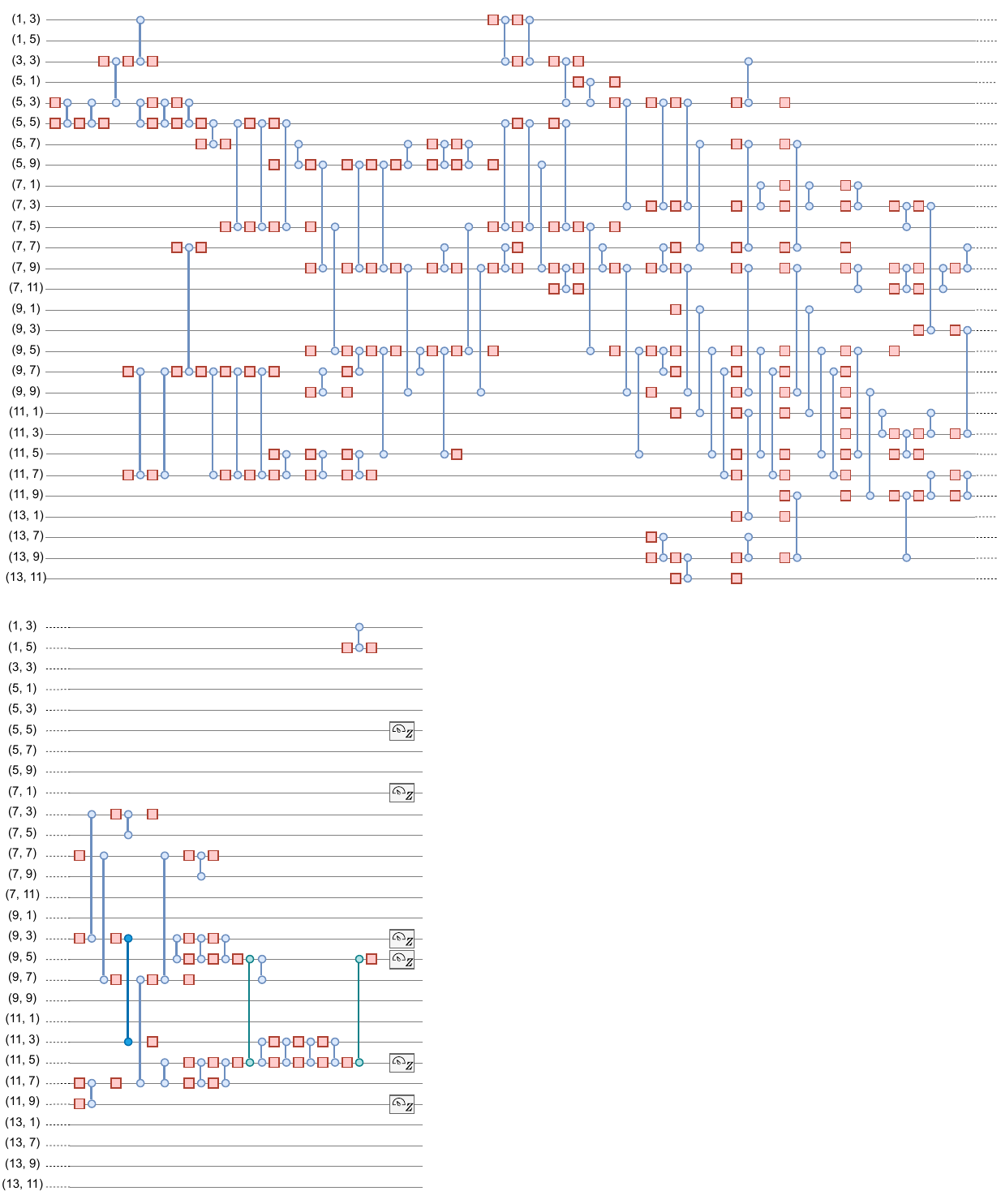}
        \caption{\textbf{The experimental quantum circuit of step 4 executed in the algebraic theorem proving.}}
        \label{fig:square4}
\end{figure*}

As an example, we present the diagrams of the experimental quantum circuits corresponding to the theoretical quantum circuit shown in Fig.~\ref{fig:square_4_theo.}. As depicted in Fig.~\ref{fig:square4}, this circuit comprises 28 physical qubits that are represented by blue dots in Fig.~\ref{fig:Perf.}\textbf{d}. On the left, we annotate these physical qubits with their coordinates in the (row index, column index) format. In total, there are 202 single-qubit gates, 116 $CP$ gates and 6 necessary $Z$ basis measurement gates. The depth of this circuit is 80. 

\subsection{Additional results}
\begin{figure*}[!htbp]
        \centering
        \includegraphics[width=\textwidth]{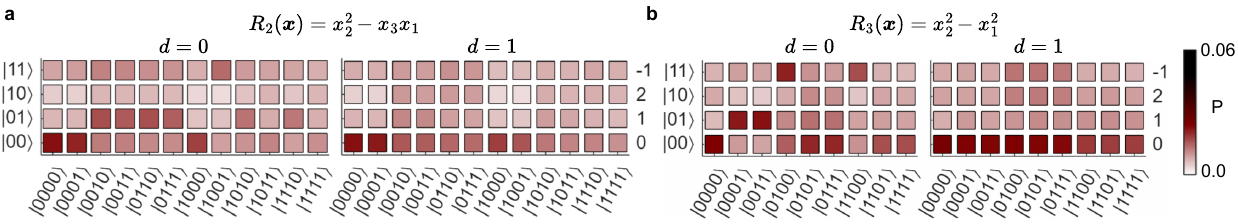}
        \caption{\textbf{The experimental point-value correspondence of step 2 (\textbf{a}) and step 3 (\textbf{b}).}
        \textbf{a}, The results of step 2 are classified into two sectors depending on the value of $d$ and the x-axis labels the joint states of point qubits $|x_1x_2x_3\rangle$. In this step, $x_1, x_3 \in \{0, 1\}$ and $x_2 \in \{00, 01, 11\}$. Colors indicate the probability obtained from 3000 runs of measurements, with each run taking 3000 shots. 
        \textbf{b}, The results of step 3 are classified into two sectors depending on the value of $d$ and the x-axis labels the joint states of point qubits $|x_1x_2\rangle$. In this step, $x_1$, $x_2 \in \{00, 01, 11\}$. Colors indicate the probability obtained from 2940 runs of measurements, with each run taking 3000 shots. 
        }             
        \label{fig:other_results}
\end{figure*}

The measurement outcomes of step 2 and step 3 are shown in Fig.~\ref{fig:other_results}\textbf{a} and \textbf{b}, respectively. In both steps, one qubit is used for the exponent register because $d$ only takes $0$ or $1$, and two qubits are used for the value register. For the point register, in step 2, depending on the specific values of each variable, we use 1, 2, 1 qubits to encode $x_1, x_2, x_3$, respectively. Similarly, in step 3, we use 2, 2 qubits to encode $x_1, x_2$, respectively. In Fig.~\ref{fig:other_results}\textbf{a} and \textbf{b}, the x- and y-axes label the joint states of the point and value registers. The coefficient value represented by each two-qubit value state is indicated along the right-hand side. 
\clearpage
\makeatother
\bibliography{Dengbib}